\documentclass[%
 reprint, twocolumn,
superscriptaddress,
 nofootinbib,
 amsmath,amssymb,
 aps, prd,
]{revtex4-2}

\usepackage[table,svgnames,dvipsnames]{xcolor}
\usepackage[mathlines]{lineno}
\usepackage[normalem]{ulem}
\usepackage{aas_macros}
\usepackage{subfigure}
\usepackage{graphicx}
\usepackage{graphics}
\usepackage{dcolumn}
\usepackage{bm}
\usepackage{cases}
\usepackage{booktabs}
\usepackage{comment}
\usepackage{multirow}
\usepackage{makecell}
\usepackage{siunitx}
\usepackage{tabularx}
\usepackage{xspace}
\usepackage{soul} 
\usepackage{fontawesome}
\usepackage{hyperref}
\hypersetup{colorlinks=true,linkcolor=blue, citecolor=blue,filecolor=blue,urlcolor=blue,}
\usepackage{adjustbox}

\usepackage{orcidlink} 
\usepackage{hanging} 

\def\be{\begin{equation}}
\def\ee{\end{equation}}

\def\ba#1\ea{\begin{align*}#1\end{align*}}

\renewcommand{\emph}[1]{\textit{#1}}

\newcommand{\lya}{Ly$\alpha$\xspace}
\newcommand{\lyaf}{Ly$\alpha$ forest\xspace}
\newcommand{\ion}[2]{#1\thinspace{}#2\xspace}

\newcommand{\hMpc}{h^{-1}\,\mathrm{Mpc}}
\newcommand{\kms}{\ensuremath{{\rm km~s^{-1}}\xspace}}

\newcommand{\Saclay}{\texttt{Saclay}\xspace}
\newcommand{\LyaCoLoRe}{\texttt{LyaCoLoRe}\xspace}

\newcommand{\QuasiLinear}{\texttt{CoLoRe-QL}\xspace}

\usepackage[nameinlink,noabbrev]{cleveref}
\crefname{equation}{Eq.}{Eqs.}
\crefname{section}{Section}{Sections}
\crefname{figure}{Figure}{Figures}
\crefname{table}{Table}{Tables}
\crefname{appendix}{Appendix}{Appendices}
\Crefname{figure}{Figure}{Figures}
\Crefname{equation}{Equation}{Equations}
\Crefname{section}{Section}{Sections}
\Crefname{table}{Table}{Tables}

\begin{document}

\title{Validation of the DESI DR2 \lya BAO analysis using synthetic datasets}


\author{L.~Casas}
\affiliation{Institut de F\'{i}sica d'Altes Energies (IFAE), The Barcelona Institute of Science and Technology, Edifici Cn, Campus UAB, 08193, Bellaterra (Barcelona), Spain}

\author{H.~K.~Herrera-Alcantar}
\affiliation{Institut d'Astrophysique de Paris. 98 bis boulevard Arago. 75014 Paris, France}
\affiliation{IRFU, CEA, Universit\'{e} Paris-Saclay, F-91191 Gif-sur-Yvette, France}

\author{J.~Chaves-Montero}
\affiliation{Institut de F\'{i}sica d’Altes Energies (IFAE), The Barcelona Institute of Science and Technology, Edifici Cn, Campus UAB, 08193, Bellaterra (Barcelona), Spain}

\author{A.~Cuceu}
\affiliation{Lawrence Berkeley National Laboratory, 1 Cyclotron Road, Berkeley, CA 94720, USA}
\affiliation{NASA Einstein Fellow}

\author{A.~Font-Ribera}
\affiliation{Institut de F\'{i}sica d’Altes Energies (IFAE), The Barcelona Institute of Science and Technology, Edifici Cn, Campus UAB, 08193, Bellaterra (Barcelona), Spain}

\author{M.~Lokken}
\affiliation{Institut de F\'{i}sica d’Altes Energies (IFAE), The Barcelona Institute of Science and Technology, Edifici Cn, Campus UAB, 08193, Bellaterra (Barcelona), Spain}

\author{M.~Abdul-Karim}
\affiliation{IRFU, CEA, Universit\'{e} Paris-Saclay, F-91191 Gif-sur-Yvette, France}

\author{C.~Ram\'irez-P\'erez}
\affiliation{Institut de F\'{i}sica d’Altes Energies (IFAE), The Barcelona Institute of Science and Technology, Edifici Cn, Campus UAB, 08193, Bellaterra (Barcelona), Spain}

\author{D.~Alonso}
\affiliation{Department of Physics, University of Oxford, Denys Wilkinson Building, Keble Road, Oxford OX1 3RH, United Kingdom}

\author{J.~Aguilar}
\affiliation{Lawrence Berkeley National Laboratory, 1 Cyclotron Road, Berkeley, CA 94720, USA}

\author{S.~Ahlen}
\affiliation{Physics Dept., Boston University, 590 Commonwealth Avenue, Boston, MA 02215, USA}

\author{U.~Andrade}
\affiliation{Leinweber Center for Theoretical Physics, University of Michigan, 450 Church Street, Ann Arbor, Michigan 48109-1040, USA}
\affiliation{University of Michigan, 500 S. State Street, Ann Arbor, MI 48109, USA}

\author{E.~Armengaud}
\affiliation{IRFU, CEA, Universit\'{e} Paris-Saclay, F-91191 Gif-sur-Yvette, France}

\author{A.~Aviles}
\affiliation{Instituto Avanzado de Cosmolog\'{\i}a A.~C., San Marcos 11 - Atenas 202. Magdalena Contreras. Ciudad de M\'{e}xico C.~P.~10720, M\'{e}xico}
\affiliation{Instituto de Ciencias F\'{\i}sicas, Universidad Nacional Aut\'onoma de M\'exico, Av. Universidad s/n, Cuernavaca, Morelos, C.~P.~62210, M\'exico}
\affiliation{Instituto de F\'{\i}sica, Universidad Nacional Aut\'{o}noma de M\'{e}xico,  Circuito de la Investigaci\'{o}n Cient\'{\i}fica, Ciudad Universitaria, Cd. de M\'{e}xico  C.~P.~04510,  M\'{e}xico}

\author{S.~Bailey}
\affiliation{Lawrence Berkeley National Laboratory, 1 Cyclotron Road, Berkeley, CA 94720, USA}

\author{S.~BenZvi}
\affiliation{Department of Physics \& Astronomy, University of Rochester, 206 Bausch and Lomb Hall, P.O. Box 270171, Rochester, NY 14627-0171, USA}

\author{D.~Bianchi}
\affiliation{Dipartimento di Fisica ``Aldo Pontremoli'', Universit\`a degli Studi di Milano, Via Celoria 16, I-20133 Milano, Italy}
\affiliation{INAF-Osservatorio Astronomico di Brera, Via Brera 28, 20122 Milano, Italy}

\author{A.~Brodzeller}
\affiliation{Lawrence Berkeley National Laboratory, 1 Cyclotron Road, Berkeley, CA 94720, USA}

\author{D.~Brooks}
\affiliation{Department of Physics \& Astronomy, University College London, Gower Street, London, WC1E 6BT, UK}

\author{R.~Canning}
\affiliation{Institute of Cosmology and Gravitation, University of Portsmouth, Dennis Sciama Building, Portsmouth, PO1 3FX, UK}

\author{A.~Carnero Rosell}
\affiliation{Departamento de Astrof\'{\i}sica, Universidad de La Laguna (ULL), E-38206, La Laguna, Tenerife, Spain}
\affiliation{Instituto de Astrof\'{\i}sica de Canarias, C/ V\'{\i}a L\'{a}ctea, s/n, E-38205 La Laguna, Tenerife, Spain}

\author{M.~Charles}
\affiliation{The Ohio State University, Columbus, 43210 OH, USA}

\author{E.~Chaussidon}
\affiliation{Lawrence Berkeley National Laboratory, 1 Cyclotron Road, Berkeley, CA 94720, USA}

\author{T.~Claybaugh}
\affiliation{Lawrence Berkeley National Laboratory, 1 Cyclotron Road, Berkeley, CA 94720, USA}

\author{K.~S.~Dawson}
\affiliation{Department of Physics and Astronomy, The University of Utah, 115 South 1400 East, Salt Lake City, UT 84112, USA}

\author{A.~de la Macorra}
\affiliation{Instituto de F\'{\i}sica, Universidad Nacional Aut\'{o}noma de M\'{e}xico,  Circuito de la Investigaci\'{o}n Cient\'{\i}fica, Ciudad Universitaria, Cd. de M\'{e}xico  C.~P.~04510,  M\'{e}xico}

\author{A.~de~Mattia}
\affiliation{IRFU, CEA, Universit\'{e} Paris-Saclay, F-91191 Gif-sur-Yvette, France}

\author{Arjun~Dey}
\affiliation{NSF NOIRLab, 950 N. Cherry Ave., Tucson, AZ 85719, USA}

\author{Biprateep~Dey}
\affiliation{Department of Astronomy \& Astrophysics, University of Toronto, Toronto, ON M5S 3H4, Canada}
\affiliation{Department of Physics \& Astronomy and Pittsburgh Particle Physics, Astrophysics, and Cosmology Center (PITT PACC), University of Pittsburgh, 3941 O'Hara Street, Pittsburgh, PA 15260, USA}

\author{Z.~Ding}
\affiliation{University of Chinese Academy of Sciences, Nanjing 211135, People's Republic of China.}

\author{P.~Doel}
\affiliation{Department of Physics \& Astronomy, University College London, Gower Street, London, WC1E 6BT, UK}

\author{D.~J.~Eisenstein}
\affiliation{Center for Astrophysics $|$ Harvard \& Smithsonian, 60 Garden Street, Cambridge, MA 02138, USA}

\author{W.~Elbers}
\affiliation{Institute for Computational Cosmology, Department of Physics, Durham University, South Road, Durham DH1 3LE, UK}

\author{S.~Ferraro}
\affiliation{Lawrence Berkeley National Laboratory, 1 Cyclotron Road, Berkeley, CA 94720, USA}
\affiliation{University of California, Berkeley, 110 Sproul Hall \#5800 Berkeley, CA 94720, USA}

\author{J.~E.~Forero-Romero}
\affiliation{Departamento de F\'isica, Universidad de los Andes, Cra. 1 No. 18A-10, Edificio Ip, CP 111711, Bogot\'a, Colombia}
\affiliation{Observatorio Astron\'omico, Universidad de los Andes, Cra. 1 No. 18A-10, Edificio H, CP 111711 Bogot\'a, Colombia}

\author{C.~Garcia-Quintero}
\affiliation{Center for Astrophysics $|$ Harvard \& Smithsonian, 60 Garden Street, Cambridge, MA 02138, USA}
\affiliation{NASA Einstein Fellow}

\author{Lehman~H.~Garrison}
\affiliation{Center for Computational Astrophysics, Flatiron Institute, 162 5\textsuperscript{th} Avenue, New York, NY 10010, USA}
\affiliation{Scientific Computing Core, Flatiron Institute, 162 5\textsuperscript{th} Avenue, New York, NY 10010, USA}

\author{E.~Gaztañaga}
\affiliation{Institut d'Estudis Espacials de Catalunya (IEEC), c/ Esteve Terradas 1, Edifici RDIT, Campus PMT-UPC, 08860 Castelldefels, Spain}
\affiliation{Institute of Cosmology and Gravitation, University of Portsmouth, Dennis Sciama Building, Portsmouth, PO1 3FX, UK}
\affiliation{Institute of Space Sciences, ICE-CSIC, Campus UAB, Carrer de Can Magrans s/n, 08913 Bellaterra, Barcelona, Spain}

\author{H.~Gil-Mar\'in}
\affiliation{Departament de F\'{\i}sica Qu\`{a}ntica i Astrof\'{\i}sica, Universitat de Barcelona, Mart\'{\i} i Franqu\`{e}s 1, E08028 Barcelona, Spain}
\affiliation{Institut d'Estudis Espacials de Catalunya (IEEC), c/ Esteve Terradas 1, Edifici RDIT, Campus PMT-UPC, 08860 Castelldefels, Spain}
\affiliation{Institut de Ci\`encies del Cosmos (ICCUB), Universitat de Barcelona (UB), c. Mart\'i i Franqu\`es, 1, 08028 Barcelona, Spain.}

\author{S.~Gontcho A Gontcho}
\affiliation{Lawrence Berkeley National Laboratory, 1 Cyclotron Road, Berkeley, CA 94720, USA}

\author{A.~X.~Gonzalez-Morales}
\affiliation{Departamento de F\'{\i}sica, DCI-Campus Le\'{o}n, Universidad de Guanajuato, Loma del Bosque 103, Le\'{o}n, Guanajuato C.~P.~37150, M\'{e}xico}

\author{C.~Gordon}
\affiliation{Institut de F\'{i}sica d’Altes Energies (IFAE), The Barcelona Institute of Science and Technology, Edifici Cn, Campus UAB, 08193, Bellaterra (Barcelona), Spain}

\author{G.~Gutierrez}
\affiliation{Fermi National Accelerator Laboratory, PO Box 500, Batavia, IL 60510, USA}

\author{J.~Guy}
\affiliation{Lawrence Berkeley National Laboratory, 1 Cyclotron Road, Berkeley, CA 94720, USA}

\author{M.~Herbold}
\affiliation{The Ohio State University, Columbus, 43210 OH, USA}

\author{K.~Honscheid}
\affiliation{Center for Cosmology and AstroParticle Physics, The Ohio State University, 191 West Woodruff Avenue, Columbus, OH 43210, USA}
\affiliation{Department of Physics, The Ohio State University, 191 West Woodruff Avenue, Columbus, OH 43210, USA}
\affiliation{The Ohio State University, Columbus, 43210 OH, USA}

\author{C.~Howlett}
\affiliation{School of Mathematics and Physics, University of Queensland, Brisbane, QLD 4072, Australia}

\author{D.~Huterer}
\affiliation{Department of Physics, University of Michigan, 450 Church Street, Ann Arbor, MI 48109, USA}
\affiliation{University of Michigan, 500 S. State Street, Ann Arbor, MI 48109, USA}

\author{M.~Ishak}
\affiliation{Department of Physics, The University of Texas at Dallas, 800 W. Campbell Rd., Richardson, TX 75080, USA}

\author{S.~Juneau}
\affiliation{NSF NOIRLab, 950 N. Cherry Ave., Tucson, AZ 85719, USA}

\author{R.~Kehoe}
\affiliation{Department of Physics, Southern Methodist University, 3215 Daniel Avenue, Dallas, TX 75275, USA}

\author{D.~Kirkby}
\affiliation{Department of Physics and Astronomy, University of California, Irvine, 92697, USA}

\author{T.~Kisner}
\affiliation{Lawrence Berkeley National Laboratory, 1 Cyclotron Road, Berkeley, CA 94720, USA}

\author{A.~Kremin}
\affiliation{Lawrence Berkeley National Laboratory, 1 Cyclotron Road, Berkeley, CA 94720, USA}

\author{O.~Lahav}
\affiliation{Department of Physics \& Astronomy, University College London, Gower Street, London, WC1E 6BT, UK}

\author{M.~Landriau}
\affiliation{Lawrence Berkeley National Laboratory, 1 Cyclotron Road, Berkeley, CA 94720, USA}

\author{J.M.~Le~Goff}
\affiliation{IRFU, CEA, Universit\'{e} Paris-Saclay, F-91191 Gif-sur-Yvette, France}

\author{L.~Le~Guillou}
\affiliation{Sorbonne Universit\'{e}, CNRS/IN2P3, Laboratoire de Physique Nucl\'{e}aire et de Hautes Energies (LPNHE), FR-75005 Paris, France}

\author{A.~Leauthaud}
\affiliation{Department of Astronomy and Astrophysics, UCO/Lick Observatory, University of California, 1156 High Street, Santa Cruz, CA 95064, USA}
\affiliation{Department of Astronomy and Astrophysics, University of California, Santa Cruz, 1156 High Street, Santa Cruz, CA 95065, USA}

\author{M.~E.~Levi}
\affiliation{Lawrence Berkeley National Laboratory, 1 Cyclotron Road, Berkeley, CA 94720, USA}

\author{Q.~Li}
\affiliation{Department of Physics and Astronomy, The University of Utah, 115 South 1400 East, Salt Lake City, UT 84112, USA}

\author{M.~Manera}
\affiliation{Departament de F\'{i}sica, Serra H\'{u}nter, Universitat Aut\`{o}noma de Barcelona, 08193 Bellaterra (Barcelona), Spain}
\affiliation{Institut de F\'{i}sica d’Altes Energies (IFAE), The Barcelona Institute of Science and Technology, Edifici Cn, Campus UAB, 08193, Bellaterra (Barcelona), Spain}

\author{P.~Martini}
\affiliation{Center for Cosmology and AstroParticle Physics, The Ohio State University, 191 West Woodruff Avenue, Columbus, OH 43210, USA}
\affiliation{Department of Astronomy, The Ohio State University, 4055 McPherson Laboratory, 140 W 18th Avenue, Columbus, OH 43210, USA}
\affiliation{The Ohio State University, Columbus, 43210 OH, USA}

\author{A.~Meisner}
\affiliation{NSF NOIRLab, 950 N. Cherry Ave., Tucson, AZ 85719, USA}

\author{J.~Mena-Fern\'andez}
\affiliation{Laboratoire de Physique Subatomique et de Cosmologie, 53 Avenue des Martyrs, 38000 Grenoble, France}

\author{R.~Miquel}
\affiliation{Instituci\'{o} Catalana de Recerca i Estudis Avan\c{c}ats, Passeig de Llu\'{\i}s Companys, 23, 08010 Barcelona, Spain}
\affiliation{Institut de F\'{i}sica d’Altes Energies (IFAE), The Barcelona Institute of Science and Technology, Edifici Cn, Campus UAB, 08193, Bellaterra (Barcelona), Spain}

\author{J.~Moustakas}
\affiliation{Department of Physics and Astronomy, Siena College, 515 Loudon Road, Loudonville, NY 12211, USA}

\author{D.~Mu\~noz Santos}
\affiliation{Aix Marseille Univ, CNRS, CNES, LAM, Marseille, France}

\author{A.~D.~Myers}
\affiliation{Department of Physics \& Astronomy, University  of Wyoming, 1000 E. University, Dept.~3905, Laramie, WY 82071, USA}

\author{S.~Nadathur}
\affiliation{Institute of Cosmology and Gravitation, University of Portsmouth, Dennis Sciama Building, Portsmouth, PO1 3FX, UK}

\author{L.~Napolitano}
\affiliation{Department of Physics \& Astronomy, University  of Wyoming, 1000 E. University, Dept.~3905, Laramie, WY 82071, USA}

\author{G.~Niz}
\affiliation{Departamento de F\'{\i}sica, DCI-Campus Le\'{o}n, Universidad de Guanajuato, Loma del Bosque 103, Le\'{o}n, Guanajuato C.~P.~37150, M\'{e}xico}
\affiliation{Instituto Avanzado de Cosmolog\'{\i}a A.~C., San Marcos 11 - Atenas 202. Magdalena Contreras. Ciudad de M\'{e}xico C.~P.~10720, M\'{e}xico}

\author{H.~E.~Noriega}
\affiliation{Instituto de Ciencias F\'{\i}sicas, Universidad Nacional Aut\'onoma de M\'exico, Av. Universidad s/n, Cuernavaca, Morelos, C.~P.~62210, M\'exico}
\affiliation{Instituto de F\'{\i}sica, Universidad Nacional Aut\'{o}noma de M\'{e}xico,  Circuito de la Investigaci\'{o}n Cient\'{\i}fica, Ciudad Universitaria, Cd. de M\'{e}xico  C.~P.~04510,  M\'{e}xico}

\author{E.~Paillas}
\affiliation{Steward Observatory, University of Arizona, 933 N. Cherry Avenue, Tucson, AZ 85721, USA}
\affiliation{Instituto de Estudios Astrof\'isicos, Facultad de Ingenier\'ia y Ciencias, Universidad Diego Portales, Av. Ej\'ercito Libertador 441, Santiago, Chile}

\author{N.~Palanque-Delabrouille}
\affiliation{IRFU, CEA, Universit\'{e} Paris-Saclay, F-91191 Gif-sur-Yvette, France}
\affiliation{Lawrence Berkeley National Laboratory, 1 Cyclotron Road, Berkeley, CA 94720, USA}

\author{W.~J.~Percival}
\affiliation{Department of Physics and Astronomy, University of Waterloo, 200 University Ave W, Waterloo, ON N2L 3G1, Canada}
\affiliation{Perimeter Institute for Theoretical Physics, 31 Caroline St. North, Waterloo, ON N2L 2Y5, Canada}
\affiliation{Waterloo Centre for Astrophysics, University of Waterloo, 200 University Ave W, Waterloo, ON N2L 3G1, Canada}

\author{Matthew~M.~Pieri}
\affiliation{Aix Marseille Univ, CNRS, CNES, LAM, Marseille, France}

\author{C.~Poppett}
\affiliation{Lawrence Berkeley National Laboratory, 1 Cyclotron Road, Berkeley, CA 94720, USA}
\affiliation{Space Sciences Laboratory, University of California, Berkeley, 7 Gauss Way, Berkeley, CA  94720, USA}
\affiliation{University of California, Berkeley, 110 Sproul Hall \#5800 Berkeley, CA 94720, USA}

\author{F.~Prada}
\affiliation{Instituto de Astrof\'{i}sica de Andaluc\'{i}a (CSIC), Glorieta de la Astronom\'{i}a, s/n, E-18008 Granada, Spain}

\author{I.~P\'erez-R\`afols}
\affiliation{Departament de F\'isica, EEBE, Universitat Polit\`ecnica de Catalunya, c/Eduard Maristany 10, 08930 Barcelona, Spain}

\author{C.~Ravoux}
\affiliation{Universit\'{e} Clermont-Auvergne, CNRS, LPCA, 63000 Clermont-Ferrand, France}

\author{G.~Rossi}
\affiliation{Department of Physics and Astronomy, Sejong University, 209 Neungdong-ro, Gwangjin-gu, Seoul 05006, Republic of Korea}

\author{E.~Sanchez}
\affiliation{CIEMAT, Avenida Complutense 40, E-28040 Madrid, Spain}

\author{D.~Schlegel}
\affiliation{Lawrence Berkeley National Laboratory, 1 Cyclotron Road, Berkeley, CA 94720, USA}

\author{M.~Schubnell}
\affiliation{Department of Physics, University of Michigan, 450 Church Street, Ann Arbor, MI 48109, USA}
\affiliation{University of Michigan, 500 S. State Street, Ann Arbor, MI 48109, USA}

\author{H.~Seo}
\affiliation{Department of Physics \& Astronomy, Ohio University, 139 University Terrace, Athens, OH 45701, USA}

\author{F.~Sinigaglia}
\affiliation{Departamento de Astrof\'{\i}sica, Universidad de La Laguna (ULL), E-38206, La Laguna, Tenerife, Spain}
\affiliation{Instituto de Astrof\'{\i}sica de Canarias, C/ V\'{\i}a L\'{a}ctea, s/n, E-38205 La Laguna, Tenerife, Spain}

\author{D.~Sprayberry}
\affiliation{NSF NOIRLab, 950 N. Cherry Ave., Tucson, AZ 85719, USA}

\author{T.~Tan}
\affiliation{IRFU, CEA, Universit\'{e} Paris-Saclay, F-91191 Gif-sur-Yvette, France}

\author{G.~Tarl\'{e}}
\affiliation{University of Michigan, 500 S. State Street, Ann Arbor, MI 48109, USA}

\author{P.~Taylor}
\affiliation{The Ohio State University, Columbus, 43210 OH, USA}

\author{W.~Turner}
\affiliation{Center for Cosmology and AstroParticle Physics, The Ohio State University, 191 West Woodruff Avenue, Columbus, OH 43210, USA}
\affiliation{Department of Astronomy, The Ohio State University, 4055 McPherson Laboratory, 140 W 18th Avenue, Columbus, OH 43210, USA}
\affiliation{The Ohio State University, Columbus, 43210 OH, USA}

\author{M.~Vargas-Maga\~na}
\affiliation{Instituto de F\'{\i}sica, Universidad Nacional Aut\'{o}noma de M\'{e}xico,  Circuito de la Investigaci\'{o}n Cient\'{\i}fica, Ciudad Universitaria, Cd. de M\'{e}xico  C.~P.~04510,  M\'{e}xico}

\author{M.~Walther}
\affiliation{Excellence Cluster ORIGINS, Boltzmannstrasse 2, D-85748 Garching, Germany}
\affiliation{University Observatory, Faculty of Physics, Ludwig-Maximilians-Universit\"{a}t, Scheinerstr. 1, 81677 M\"{u}nchen, Germany}

\author{B.~A.~Weaver}
\affiliation{NSF NOIRLab, 950 N. Cherry Ave., Tucson, AZ 85719, USA}

\author{M.~Wolfson}
\affiliation{The Ohio State University, Columbus, 43210 OH, USA}

\author{C.~Yèche}
\affiliation{IRFU, CEA, Universit\'{e} Paris-Saclay, F-91191 Gif-sur-Yvette, France}

\author{P.~Zarrouk}
\affiliation{Sorbonne Universit\'{e}, CNRS/IN2P3, Laboratoire de Physique Nucl\'{e}aire et de Hautes Energies (LPNHE), FR-75005 Paris, France}

\author{R.~Zhou}
\affiliation{Lawrence Berkeley National Laboratory, 1 Cyclotron Road, Berkeley, CA 94720, USA}

\collaboration{DESI Collaboration}

\begin{abstract}
The second data release (DR2) of the Dark Energy Spectroscopic Instrument (DESI), containing data from the first three years of observations, doubles the number of Lyman-$\alpha$ (Ly$\alpha$\xspace) forest spectra in DR1 and it provides the largest dataset of its kind. 
To ensure a robust validation of the Baryonic Acoustic Oscillation (BAO) analysis using Ly$\alpha$\xspace forests, we have made significant updates compared to DR1 to both the mocks and the analysis framework used in the validation. 
In particular, we present \texttt{CoLoRe-QL}, a new set of Ly$\alpha$\xspace mocks that use a quasi-linear input power spectrum to incorporate the non-linear broadening of the BAO peak. 
We have also increased the number of realisations used in the validation to 400, compared to the 150 realisations used in DR1. 
Finally, we present a detailed study of the impact of quasar redshift errors on the BAO measurement, and we compare different strategies to mask Damped Lyman-$\alpha$ Absorbers (DLAs) in our spectra.
The BAO measurement from the Ly$\alpha$\xspace dataset of DESI DR2 is presented in a companion publication.
\end{abstract}

\maketitle

\tableofcontents

\section{Introduction} 
\label{sec:introduction}

One of the central questions in cosmology is the cause behind the accelerated expansion of the Universe. Among the most powerful observables for investigating this phenomenon are baryon acoustic oscillations (BAO), fluctuations in the matter density caused by acoustic density waves in the early universe. First measured in the distribution of galaxies twenty years ago \cite{Cole2005, Eisenstein2005}, BAO serves as a standard ruler for measuring cosmic expansion.

Measuring BAO at $z>2$ using galaxies as tracers is challenging, as obtaining a sufficient number of redshifts for distant, faint galaxies is highly time-consuming. However, the Lyman-$\alpha$ (\lya) forest in the spectra of distant quasars provides a powerful alternative. This observable consists of absorption lines in the spectra of high-redshift quasars caused by neutral hydrogen in the intergalactic medium. Consequently, the \lya forest traces the distribution of matter in the Universe, enabling the measurement of BAO at higher redshifts than those accessible by galaxy surveys. The Baryon Oscillation Spectroscopic Survey (BOSS, \cite{Dawson2013}) was the first to carry out these measurements from the auto-correlation of the \lya forest \cite{Busca2013, Slosar2013, Kirkby2013} and its cross-correlation with quasars \cite{FontRibera2014}.

The Dark Energy Spectroscopic Instrument (DESI) represents a major step forward in the precise measurement of BAO from the \lya forest \citep{Snowmass2013.Levi, DESI2016a.Science}. DESI is a robotic, fiber-fed, highly multiplexed spectroscopic surveyor that operates on the Mayall 4-meter telescope at Kitt Peak National Observatory \cite{DESI2022.KP1.Instr} that can obtain simultaneous spectra of almost 5000 objects \citep{DESI2016b.Instr, FocalPlane.Silber.2023, Corrector.Miller.2023, FiberSystem.Poppett.2024} using complex planning \cite{SurveyOps.Schlafly.2023} and reduction pipelines \cite{Spectro.Pipeline.Guy.2023}. 
The DESI instrument and spectroscopic pipelines were tested extensively during a period of survey validation before the start of the main survey \citep{DESI2023a.KP1.SV,DESI2023b.KP1.EDR}.
The first year of DESI observations (DR1, \citep{DESI2024.I.DR1}) yielded the most precise measurement of BAO up to that time from the \lya forest using over 420\,000 \lya forest spectra and 700\,000 quasars \cite{DESI2024.IV.KP6}. 
This dataset also led to precise BAO \citep{DESI2024.III.KP4} and full-shape \citep{DESI2024.V.KP5} measurements from the clustering of galaxies and quasars \citep{DESI2024.II.KP3}. 
In combination with external probes, these measurements from DESI DR1 resulted in one of the tightest constraints on dynamical dark energy, and on the sum of the neutrino masses \citep{DESI2024.VI.KP7A, DESI2024.VII.KP7B}. 

The first three years of DESI observations (DR2, \cite{DESI.DR2.DR2}) include over 820,000 \lya forest spectra and 1.2 million quasars at $z>1.77$, almost doubling the dataset in DR1. The goal of this work is to use synthetic realizations of DESI data (mock catalogs) to validate the \lya BAO analysis of the second data release of DESI \citep{DESI.DR2.BAO.lya}. Mock datasets provide a controlled environment where all possible sources of systematics are known, since these mock datasets provide a controlled environment to study systematic effects. Therefore they are essential for testing the analysis pipeline and assessing the impact of potential systematic errors on cosmological constraints. 

Compared to the validation of the DR1 measurement presented in \citep{KP6s6-Cuceu}, we have introduced several improvements to both the mocks themselves and to their analysis. These include refining the accuracy of quasar small-scale clustering, incorporating non-linear broadening of the BAO peak, improving the treatment of redshift errors and Damped Lyman-$\alpha$ Absorbers (DLAs), and increasing the number of mocks from 150 to 400.
Since the DESI DR2 \lya BAO analysis is limited to scales larger than $30~\hMpc$, expected to be fairly well described by linear theory \cite{Givans2022}, we use relatively simple lognormal mocks and ignore the impact of non-linearities beyond the broadening of the BAO peak.

Our results are part of a comprehensive set of DESI DR2 BAO measurements from the clustering of galaxies, quasars, and the \lya forest. The companion paper \citep{DESI.DR2.BAO.cosmo} presents the BAO measurements from galaxies and quasars at $z < 2$ and the cosmological interpretation of all BAO measurements.

The outline of the paper is as follows. \Cref{sec:mocks} provides an overview of the mock datasets used in DR1 and DR2, focusing on the updates made for the validation of DR2.
\Cref{sec:methodology} describes the analysis methodology applied to the mocks, outlining the steps involved in extracting and interpreting the BAO signal. 
The main results are presented in \cref{sec:results}, and in \cref{sec:discussion} we discuss the small bias present in BAO measurements when we introduce redshift errors before continuum fitting.
Finally, \cref{sec:conclusions} summarizes the findings and presents the conclusions of this work.

\section{Mocks} 
\label{sec:mocks}

This section outlines the process used to create the mocks for validating the \lya BAO analysis in DR2, that can be divided into two stages. In the first one, explained in \cref{sec:transmission}, we simulate the distribution of quasars and extract the \lya transmitted flux fraction along the line-of-sight to each quasar, which we usually refer to as transmission skewers. 
The second stage transforms these transmission skewers into realistic DESI spectra, as detailed in \cref{ss:quickquasars}.

\subsection{Simulations of the Universe at $z>2$}
\label{sec:transmission}

In DR1, two different types of mocks were used to simulate the distribution of quasars and the fluctuations in the \lya forest: 100 realizations of \LyaCoLoRe mocks \cite{Farr2020_LyaCoLoRe} and 50 realizations of \Saclay mocks \cite{Etourneau2023}.
For the analysis validation of DR2 we have used 300 realizations of \QuasiLinear mocks (improved version of \LyaCoLoRe mocks used in DR1, described in \cref{ss:colore_updates}) and 100 realizations of the \Saclay mocks.
Both sets of mocks are based on local transformations of Gaussian random fields, and do not include higher order correlations arising from the non-linear evolution of the density field. However, they have different approaches to populate the simulated boxes with quasars, and to add redshift-space distortions in the \lya forest.

\subsubsection{Previous mocks already used in DESI DR1}

The first step to generate a \LyaCoLoRe or \QuasiLinear mock is to produce 
a 3-dimensional 10 $\rm{Gpc}\ h^{-1}$ Gaussian random field using an input power spectrum corresponding to the linear power spectrum of density fluctuations at $z=0$, and extrapolate it back in time to generate an all-sky light-cone reaching $z=3.8$.
We then use a \textit{biasing model} to translate this field into fluctuations in the density of quasars, that we Poisson sample to obtain our catalog of quasar positions. 
This is done using the \texttt{CoLoRe} package\footnote{\url{https://github.com/damonge/CoLoRe}}, described in \cite{Ramirez2022}, that is also used to compute the Gaussian \textit{skewers} of both densities and line-of-sight velocities from the center of the box (the observer) to each of the quasar positions. 
We then use the package \texttt{LyaCoLoRe}\footnote{\url{https://github.com/igmhub/LyaCoLoRe}}, described in \cite{Farr2020_LyaCoLoRe}, to translate these Gaussian skewers into the redshift-space \lya optical depth, from which we compute the transmitted flux fraction $F=e^{-\tau}$. This process involves several steps. First, given that the cells of the initial field are $\simeq 2.4 \hMpc$, in order to reproduce the 1D fluctuations in the \lya forest we need to add extra small-scale power to the Gaussian skewers. This is followed by a lognormal transformation in order to obtain positive densities. Next, we apply the Fluctuating Gunn-Peterson Approximation (FGPA, \cite{Croft1998}) to compute the real-space optical depth. We then use the line-of-sight velocities to shift the absorption and obtain the redshift-space optical depth. Finally, we compute the transmitted flux fraction.
We refer the reader to \cite{Farr2020_LyaCoLoRe} for more details on these mocks.

\Saclay mocks, on the other hand, were created using the \texttt{SaclayMocks}\footnote{\url{https://github.com/igmhub/SaclayMocks}} package. The methodology is similar to that of the \LyaCoLoRe mocks, but it presents two important differences.
Instead of using the same random field to simulate both the quasar positions and the \lya skewers, here we use two random fields that have the same random seed, but different input power spectra. 
 The input power spectrum for these mocks is chosen so that the resulting quasar power spectrum remains approximately linear, with a scale-independent bias and linear redshift-space distortions down to small scales.
The second difference is the implementation of redshift-space distortions in the \lya skewers. The velocity gradients along the lines of sight are also computed from the initial Gaussian density field. However, a modified FGPA transformation is then applied to the sum of the density and velocity gradient fields, generating directly the redshift-space optical depth. A multiplicative factor is applied to the velocity gradient to fit the measured amount of redshift space distortions. 
This implementation allows for a predictive model of the correlations down to small scales \cite{Etourneau2023}.

\subsubsection{New, Quasi-Linear mocks (\QuasiLinear)}
\label{ss:colore_updates}

As shown in figure 16 of \cite{KP6s6-Cuceu}, the BAO uncertainties obtained when analyzing the DR1 mocks were systematically smaller than the BAO uncertainties in DESI DR1.
The discrepancy could be explained by the non-linear broadening of the BAO peak, present in data, but absent in our Gaussian mocks. Non-linear evolution causes an increased broadening of the peak, making it more smeared compared to linear evolution and reducing the precision of its position measurement.
In future analyses of DESI \lya we plan to incorporate this effect by using more complex simulations based on perturbation theory.
Meanwhile, in this publication we have used a modified version of the \LyaCoLoRe mocks with broadened BAO features. Instead of using the linear power spectrum to generate the Gaussian density field, we use now as input a \textit{quasi-linear} power spectrum that is similar to the one described in \cref{sec:methodology_model} and used in the BAO fits.
In particular we use an isotropic version of \cref{eq:pksmooth}, evaluated at $k_\parallel=0$ and with $\Sigma_\perp = 3.26~\hMpc$. 
Given that the velocity fields in \texttt{CoLoRe} are derived from the Gaussian density field, and are therefore also affected by this smoothing, the final non-linear broadening of the BAO peak is somewhat anisotropic. 
In order to fully capture the anisotropic broadening of the peak one would need to use mocks with more realistic clustering, such as LPT-based mocks, something that we leave for future work.
However, as we discuss in Section \ref{sec:results}, the BAO uncertainties from the \QuasiLinear mocks are already in better agreement with the ones measured in DESI DR2, while the \Saclay mocks (that have not been updated) still show smaller BAO uncertainties. 

\begin{figure}[htbp] 
    \centering         \includegraphics[width=0.95\linewidth]{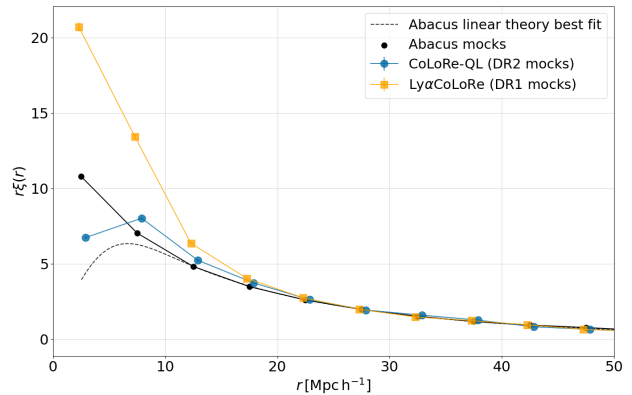}  
    \caption{Comparison of the QSO auto-correlation function measured in the \LyaCoLoRe mocks used in DR1 (yellow), the \QuasiLinear mocks presented here (blue), and the quasar catalogs from the Abacus simulation (black) at $z=2.5$. 
    The dashed black line is the best-fit linear theory to the Abacus mocks.}
    \label{fig:qso-cf} 
\end{figure}

As shown in Figure 6 of \cite{Youles2022}, the clustering of quasars in the \LyaCoLoRe mocks used to validate the analysis of eBOSS DR16 and in DESI DR1 did not reproduce well the clustering seen in the quasar catalogs from the more realistic Abacus simulations \cite{yuan2023desionepercentsurveyexploring}, which we take as our benchmark for comparison. 
In order to improve this, the \QuasiLinear mocks use a different biasing model than the exponential one used in the DR1 mocks \cite{Ramirez2022}. 
We use a linear biasing model, where the density of quasars is proportional to the (lognormal) density of matter, but modified such that quasars can only populate cells with a (lognormal) matter density threshold following:

\begin{align}
1 + \delta_Q &\propto 1 + b_Q \delta_M &\text{for}\ \delta_M > t \nonumber \\
1 + \delta_Q &= 0 &\text{for}\ \delta_M \leq t
\label{bias model}
\end{align}
where $t$ is the threshold and $\delta_Q$ and $\delta_M$ represent the quasar and matter overdensities, respectively.
The values of the linear bias and of the threshold density were tuned to match the amplitude of quasar clustering on linear scales from a preliminary, blinded measurement of quasar clustering in DESI DR1. Both parameters vary as functions of redshift, and their precise configuration details are available in the corresponding configuration file\footnote{The configuration files can be accessed at \url{https://github.com/igmhub/LyaCoLoRe/tree/colore-ql/colore-ql}}
In \cref{fig:qso-cf}, we compare the quasar clustering from \LyaCoLoRe, \QuasiLinear, and Abacus mocks. The results show that the \QuasiLinear mocks are in better agreement with the Abacus mocks down to smaller scales.

\subsection{Synthetic quasar spectra for DESI DR2}
\label{ss:quickquasars}

After generating the quasar positions and the skewers of \lya transmission described in \cref{sec:transmission}, referred to here as the \textit{raw mocks}, we use the \texttt{desisim}\footnote{\url{https://github.com/desihub/desisim}} package to emulate the characteristics of the DESI DR2 \lya quasar sample as closely as possible. 

\begin{figure}[htbp]
    \centering
    \includegraphics[width=0.49\textwidth]{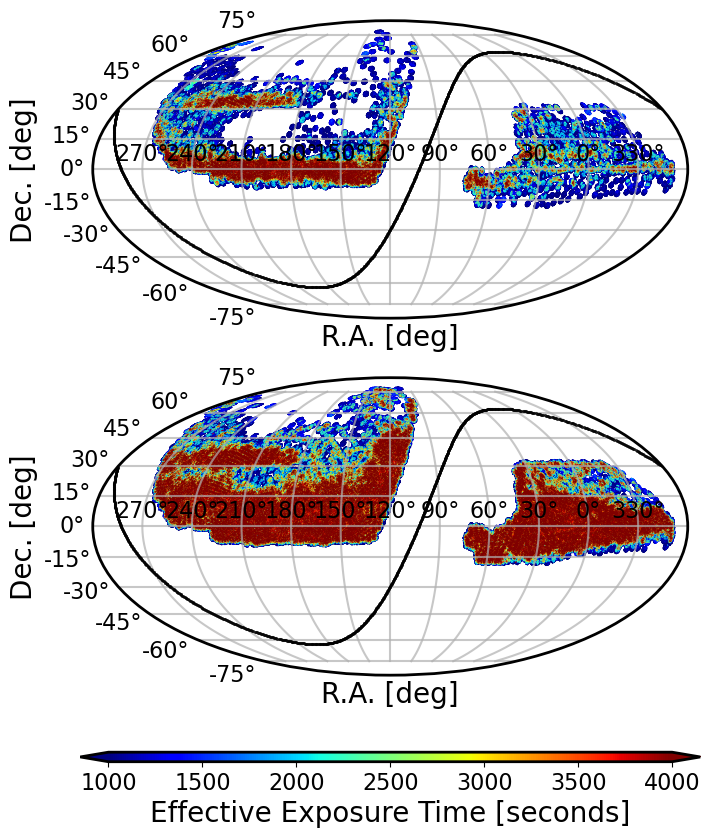}
    \caption{Comparison of the footprints in DESI DR1 (top) and DR2 (bottom). The colors show the effective exposure time. }
    \label{fig:footprint}
\end{figure}

The first step is to downsample the simulated quasar catalog to match the footprint and redshift distribution of quasars in DR2.
Next, we randomly assign an $r$-band magnitude and number of exposures to each quasar following the observed distributions, 
as described in Section 2.2 of \cite{KP6s6-Cuceu}. \Cref{fig:footprint} shows a comparison of the footprints in DR1 and DR2 of DESI. 
Since our mocks do not have gravitationally collapsed objects, we 
mimic the impact of non-linear peculiar velocities by adding random shifts to the quasar redshifts, following a Gaussian distribution with a dispersion of $\sigma_{\rm FoG} = 150~\kms$, which is the expected velocity dispersion for halos with $\sim 10^{12}h^{-1}\, M_\odot$ characteristic mass hosting quasars at $z>2.0$~\citep{eBOSS:2017Laurent}.

\begin{figure*}
\centering
\includegraphics[width=0.9\linewidth]{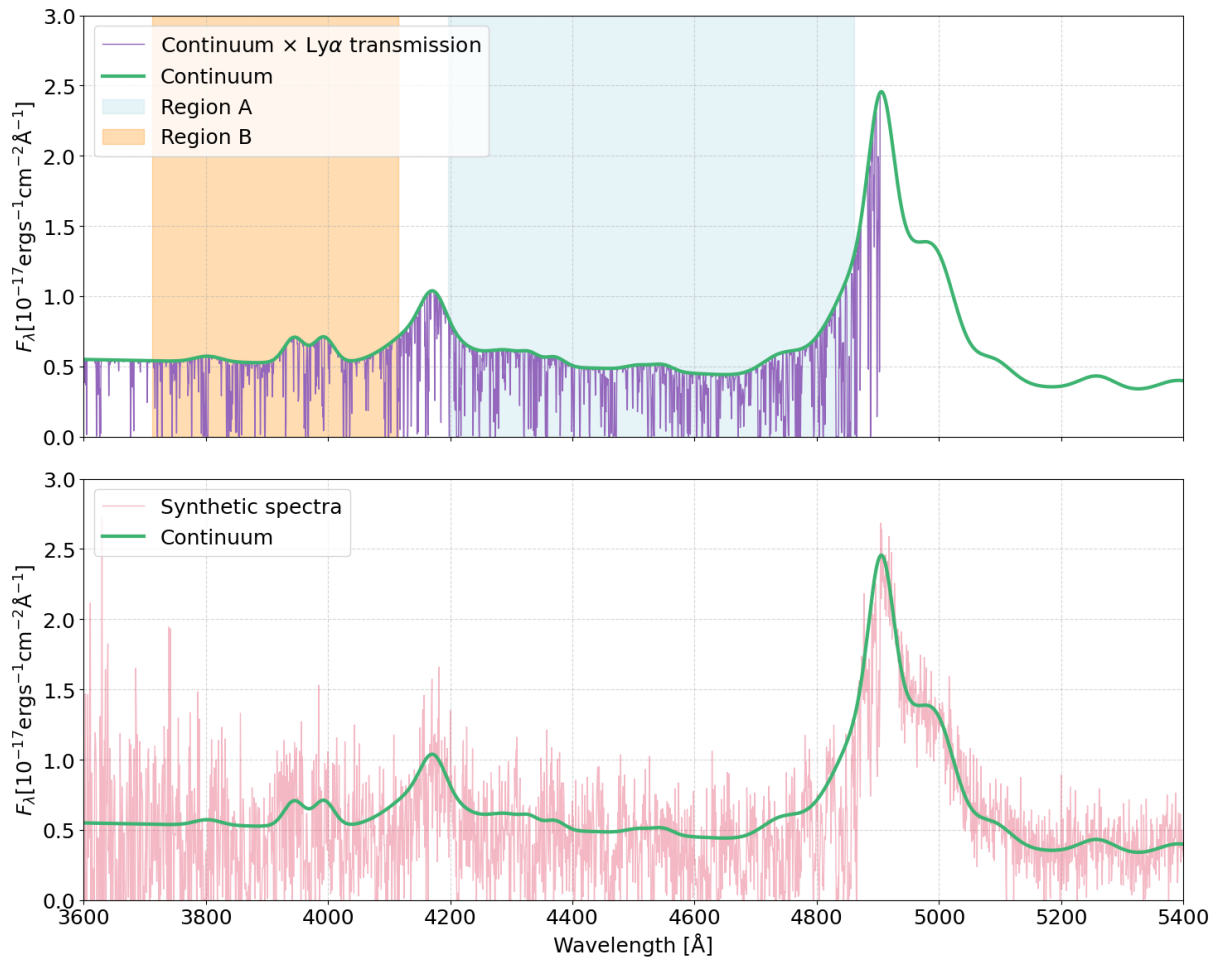}
\caption{Simulated spectra from a \QuasiLinear\ mock at $z=3.03$. Top panel: The green curve represents the quasar continuum as calculated by \texttt{quickquasars}, while the purple lines show the \lya absorption, obtained by multiplying the transmission flux fraction with the quasar continuum. The shaded regions, A (blue) and B (orange), highlight the different areas used for calculating the correlations. Bottom panel: The pink line illustrates the simulated flux after incorporating instrumental noise using \texttt{specsim}.}
\label{fig:spectra}
\end{figure*}

The second step is to generate synthetic spectra for each of the simulated quasars. For this we use the script \texttt{quickquasars}\footnote{\url{https://github.com/desihub/desisim/blob/main/py/desisim/scripts/quickquasars.py}}, described in detail in \citep{2024arXiv240100303H}.
The \lya skewers with the transmitted flux fraction a shown in the top panel of Figure \ref{fig:spectra} are modified to include various astrophysical contaminants, such as High Column Density systems (HCDs), Broad Absorption Lines (BALs) and other transition lines (metals).
These are then multiplied by random (unabsorbed) quasar spectra (quasar continua), which are generated following the \texttt{SIMQSO}\footnote{\url{https://github.com/imcgreer/simqso}} package \citep{simqso:2021}. 
Finally, we use the \texttt{specsim}\footnote{\url{https://github.com/desihub/specsim}} package \citep{Kirkby:2016} to add instrumental noise representative of DESI during nominal dark-time conditions
.
An example of a mock spectrum after adding instrumental noise is shown on the bottom panel of Figure \ref{fig:spectra}.
For further details on the synthetic spectra generation procedure, we refer the reader to Sections 2 and 3 of \cite{2024arXiv240100303H}.

As discussed in \cite{DESI.DR2.BAO.lya}, approximately 15\% of the observed quasars in DR2 were originally targeted as Emission Line Galaxies (ELGs).
These have a different redshift-magnitude distribution and only have one observation (\lya quasars in DESI can have up to four observations). 
These quasars were not included in the DR1 mocks presented in \cite{KP6s6-Cuceu}. 
In the DR2 mocks we have included these extra quasars, using similar recipes to the ones used to generate the quasar targets, and ignoring the fact that quasars targeted as ELGs could have different spectral properties than those targeted as quasars. 

After generating the mock spectra, we create in post-process a quasar catalog where we additionally include a random error on the reported redshift that emulates the statistical uncertainties from quasar redshift estimation algorithms such as \texttt{redrock}~\citep{Redrock.Bailey.2024,RedrockQSO.Brodzeller.2023} or \texttt{QuasarNET}~\citep{Busca18}. 
We add these errors following a Gaussian distribution with a dispersion of $\sigma_z = 400~\kms$, based on the measurements from \cite{KP6s4-Bault}.
Redshift errors cause a contamination in the measured correlations that was first identified in \cite{Youles2022}, and their impact on the measurement of BAO was partly discussed in section 4.4 of the paper describing the validation of the DR1 \lya analysis~\citep{KP6s6-Cuceu}. 
In this work, we include these redshift errors in the baseline configuration of the analysis. We discuss the results and propose an strategy to mitigate their effect in \cref{sec:discussion}.

\section{Methodology} 
\label{sec:methodology}

This work aims to validate the BAO analysis of \lyaf measurements from DESI DR2 using the mocks described in \cref{sec:mocks}. In this section, we outline the entire process from simulated spectra to BAO measurements. We describe our method for extracting \lyaf catalogs in \cref{sec:methodology_forest}, computing the \lya flux overdensity field in \cref{sec:methodology_deltas}, estimating \lya correlation functions in \cref{sec:methodology_correlation}, and modeling these correlations in \cref{sec:methodology_model}. Finally, we explain how the BAO parameters are extracted by fitting the model to the measured correlation functions in \cref{sec:methodology_fitting}. The first three steps are implemented using the publicly available \texttt{picca}\footnote{\url{https://github.com/igmhub/picca}},
while the modeling and fitting are conducted with the \texttt{Vega}\footnote{\url{https://github.com/andreicuceu/vega}} package. 
The methodology followed here is similar to the one used in the analysis of the DR1 mocks \cite{KP6s6-Cuceu}, but we have introduced a couple of changes that we highlight below.


\subsection{The \lyaf catalog}
\label{sec:methodology_forest}

In this section, we summarize the procedure for deriving a \lyaf catalog from simulated quasar spectra. We begin by applying two observed-frame wavelength cuts to the spectra. The lower limit, $\lambda_\mathrm{min} = 3600\, \mathrm{\AA}$, corresponds to the minimum wavelength detected by the blue arm of the DESI spectrographs, while the upper limit, $\lambda_\mathrm{max} = 5772 \, \mathrm{\AA}$, marks the midpoint of the overlap region between the blue and red arms of the spectrographs.

We continue by extracting the \lyaf from two distinct regions of the spectra. As shown in \cref{fig:spectra}, region A spans the rest-frame wavelength range $\lambda_\mathrm{RF}[\mathrm{\AA}]\in[1040, 1205]$, which is delimited by \lya and Ly$\beta$ lines, while region B covers $\lambda_\mathrm{RF}[\mathrm{\AA}]\in[920, 1020]$. The rationale for using two regions is that region A contains absorption exclusively from the first line of the Lyman series, \lya, while region B also includes higher-order absorption lines such as Ly$\beta$ and Ly$\gamma$. 
Although both regions are affected by absorption from other transition lines (metals), we assume all absorption is due to \lya when measuring the correlation functions (\cref{sec:methodology_correlation}) and later model their impact on these (\cref{sec:methodology_model}).

We use \textit{\lya pixels} to refer to those pixels in the rest-frame wavelength regions defined above, and we
refer to the collection of all \lya pixels within a quasar spectrum as a forest. 
The combination of the observed- and rest-frame selection criteria
constrain the redshift range of quasars contributing \lya pixels in the region A to $z \in [2.1, 4.4]$ and in the region B to $z \in [2.6, 5.1]$. The number of quasars contributing to regions A and B in the mocks is approximately 800\,000 and 360\,000, respectively.

Before extracting the \lya overdensity field, we apply corrections to mitigate the impact of two astrophysical contaminants: damped \lya systems (DLAs) and broad absorption line quasars (BALs). We first identify pixels associated with DLAs detected at a signal-to-noise ratio greater than 2 and with column densities $N_{\rm HI} > 10^{\mathrm{20.3}}{\rm cm}^{-\mathrm{2}}$. The DLA detection algorithm achieves a completeness of 75\% for such systems \cite{Y3.lya-s2.Brodzeller.2025}. 
To replicate the impact of DLAs in observations, we randomly mask the same fraction of DLAs that satisfy these criteria in the mocks. Specifically, we mask all DLA pixels where the transmitted flux decreases by 20\% or more and correct the transmitted flux of the remaining DLA pixels using a Voigt profile \cite{Noterdaeme2012} (see also \cref{app:dla_masking}). For the validation of the DR1 measurement using mocks, we assumed 100\% completeness for the DLA finder, thereby masking all DLAs in the mocks \cite{KP6s6-Cuceu}.

We mask the expected locations of all potential BAL features, regardless of whether the absorption is apparent \cite{Ennesser2022}. These features include \lya, \ion{N}{IV}, \ion{C}{III}, \ion{Si}{IV}, and \ion{P}{V} in region A, and \ion{O}{VI}, \ion{O}{I}, Ly$\beta$, Ly$\gamma$, \ion{N}{III}, and Ly$\delta$ in region B. Finally, we discard forests with fewer than 150 pixels, which is a threshold required by the continuum fitting procedure (see next section).


\subsection{The flux transmission field}
\label{sec:methodology_deltas}

In this section, we provide a brief overview of our method for measuring the \lya flux transmission field. We follow the same approach used for analyzing DR1 measurements; see \cite{DESI2024.IV.KP6, KP6s6-Cuceu}, for more details. 

We begin by calculating the \lya flux overdensity field, given by
\begin{equation}\label{eq:delta}
    \delta_q = \frac{f_q(\lambda)}{\bar{F}(z) C_q(\lambda)} - 1,
\end{equation}
where $f_q$ is the observed flux density for a quasar $q$, $C_q$ is the unabsorbed flux density (also referred to as the quasar continuum), $\bar{F}$ is the mean transmission of the intergalactic medium (IGM) at the absorber redshift $z = \lambda/\lambda_\alpha - 1$, and $\lambda_\alpha$ is the rest-frame wavelength of \lya.

Measuring the flux overdensity field requires estimating the product $\bar{F}(z) C_q(\lambda)$ for each quasar. This step is known as continuum fitting, which is described in detail in \cite{dMdB2020, 2023MNRAS.tmp.3626R}. Here, we provide a brief overview of the process, noting that we perform it separately for regions A and B.
First, we approximate the product $\bar{F}(z) C_q(\lambda)$ for each quasar as the product of $\bar{C}$, which is the same for all quasars, and a quasar-specific first-degree polynomial in $\Lambda = \log\lambda$,
\begin{equation}\label{eq:delta_approx}
    E_q^X = \bar{C}^X(\lambda_{\mathrm{RF}})\left(a_q^X + b_q^X\frac{\Lambda-\Lambda_{\mathrm{min}}}{\Lambda_{\mathrm{max}}-\Lambda_{\mathrm{min}}}\right),
\end{equation}
where $X$ refers to regions A and B, $\lambda_\mathrm{RF} = \lambda/(1+z_q)$, and $\Lambda_\mathrm{max}$ and $\Lambda_\mathrm{min}$ correspond to the maximum and minimum observed-frame wavelengths of the forest for the quasar $q$.
We fit $a_q^X$ and $b_q^X$ by maximizing the likelihood function
\begin{equation}
    \ln{\mathcal L} = -0.5\sum_{i}\left[\frac{f_q(\lambda_i) - E_{q}^X(\lambda_{i})}{\sigma_q^X(\lambda_{i}, a_q^X, b_q^X)}\right]^2 - \sum_{i}\ln{\sigma^X_{q}\left(\lambda_{i}, a_q^X, b_q^X \right)},
    \label{eqn:aqbq_likelihood}
\end{equation}
where $(\sigma_q^X)^2 = \eta_{\rm pip}^X (\lambda)\sigma_{{\rm pip},\,q}^2(\lambda) + [\sigma^X_{\rm LSS}(\lambda) E_q^X(\lambda)]^2$ is the flux variance of each pixel, $\sigma_{{\rm pip}, q}^2$ is the pipeline estimate for the flux variance, $\eta_{\rm pip}^X$ is a correction factor for inaccuracies in this estimate, and $\sigma^X_{\rm LSS}$ represents the intrinsic standard deviation of fluctuations in the \lya forest.
Continuum fitting is an iterative process that begins by assuming initial values for $\bar{C}^X$, $\eta_{\rm pip}^X$, and $\sigma^X_{\rm LSS}$. The coefficients $a_q^X$ and $b_q^X$ are then computed for each region of all quasars. Next, we estimate the \lya flux overdensity field for all forests, calculate its variance, and adjust the values of $\eta_{\rm pip}^X$ and $\sigma^X_{\rm LSS}$. Finally, we measure $\bar{C}^X$ and repeat the whole process until convergence, which typically takes 5 steps.

As first discussed in \cite{Slosar2011,FontRibera2012b}, when fitting the $a_q^X$ and $b_q^X$ coefficients for each quasar we are suppressing very long wavelength fluctuations in the \lya forest, distorting the measured correlations. 
In order to make the modeling of the distortion easier, we follow \cite{Bautista2017} and we explicitly subtract the mean and first moment from each forest. This is done by applying the projection operator 
$\zeta$ to the fluctuation field where,

\begin{equation}
    \zeta^{q}_{ij} = \delta_{ij}^K - \frac{w_j}{\Sigma_kw_k}- \frac{w_j(\Lambda_i - \bar \Lambda_q)(\Lambda_j - \bar \Lambda_q)}{\Sigma_kw_k (\Lambda_k - \bar \Lambda_q)^2}
\end{equation}

and redefining the fluctuations as

\begin{equation}
    \tilde{\delta}^q(\lambda_i) = \Sigma_j \zeta^q_{ij}\delta^q(\lambda_j)
\end{equation}

where $\delta^K$ is the Kronecker delta and $\bar \Lambda_q$ is the mean of $\Lambda$ for quasar $q$.  The weights are identical to those used in the correlation measurement (see \cref{eq:lya_weight}). For simplicity we will refer to this projected density fluctuation as $\delta$ in the rest of the text.


\subsection{Measuring the correlation functions}
\label{sec:methodology_correlation}

There are six possible correlations involving quasar positions and \lya fluctuations in regions A and B. However, following recent \lya BAO analyses \citep{dMdB2020,DESI2024.IV.KP6}, we focus on those with the highest signal-to-noise ratio: the auto-correlation of \lya fluctuations in region A, $\xi^\mathrm{AA}$; the cross-correlation of \lya fluctuations between regions A and B, $\xi^\mathrm{AB}$; and the cross-correlation of quasar positions with \lya fluctuations in regions A and B, denoted as $\xi^\mathrm{QA}$ and $\xi^\mathrm{QB}$, respectively. 
In this section, we briefly outline our methodology for measuring these correlations, which is the same as the one used in the DR1 analysis and its validation. See \cite{2023JCAP...11..045G, KP6s6-Cuceu} for further details.

We compute the correlation functions on a grid of separations, $r_\parallel$ and $r_\perp$, corresponding to distances along and across the line of sight, respectively. To define these components  we adopt the angular bisector LOS convention as explained in \cite{Beutler:2018vpe}. This computation requires assuming a fiducial cosmology to convert angular and redshift differences into comoving separations
\begin{equation}
r_\parallel = [D_C(z_i) - D_C(z_j)]~\mathrm{cos}\left(\frac{\theta_{ij}}{2}\right), 
\end{equation}
\begin{equation}
r_\perp = [D_M(z_i) + D_M(z_j)]~\mathrm{sin}\left(\frac{\theta_{ij}}{2}\right),
\end{equation}
where $i$ and $j$ index pixel-pixel or pixel-quasar pairs, pixel redshifts are determined by assuming \lya absorption as $z_i=\lambda/\lambda_\alpha-1$, $\theta_{ij}$ refers to the angular separation, and $D_C$ and $D_M$ denote the comoving and angular comoving distances, respectively, which are identical in a flat Universe. We use the best-fitting flat $\Lambda$CDM model from {\it Planck} 2015 results \cite{Planck2016_cosmo} as our fiducial cosmology, the same model used to generate the \LyaCoLoRe mocks\footnote{Note that for the analysis of observational measurements we use as fiducial cosmology {\it Planck} 2018 \cite{Planck2018}.}.

The algorithms to compute the auto and the cross-correlations can be expressed in the following compact form
\begin{equation}\label{eq:autocorrcalc}
\hat{\xi}_M^{XY} = \frac{\sum_{i,j \in M} w_i^X w_j^Y \hat{\delta}_i^X \hat{\delta}_j^Y}{\sum_{i,j \in M} w_i^X w_j^Y},
\end{equation}
where $X$ and $Y$ correspond to the auto- and cross-correlation cases defined at the beginning of this section, $M$ represents a two-dimensional bin with widths $(\Delta r, \Delta r)$, and $\delta^\mathrm{Q}=1$. The weight assigned to quasar $i$ is given by $w_i^\mathrm{Q} = \left[(1+z_i)/(1+z_\mathrm{Q})\right]^{\gamma_\mathrm{Q}-1}$, where $z_\mathrm{Q}=2.25$ and $\gamma_\mathrm{Q}=1.44$ are derived from previous analyses \cite{dMdB2017}. The weights for \lya fluctuations in regions A and B are
\begin{equation} \label{eq:lya_weight}
 w_i^X = \frac{\left[(1+z_i)/(1+z_{\rm fid}) \right]^{\gamma_{\alpha}-1}}
 {\eta_{\rm pip}^X(\lambda) \sigma^2_{\mathrm{pip}, q}(\lambda) [E_q^X(\lambda)]^{-2} + \eta_\mathrm{LSS}[\sigma^X_\mathrm{LSS}(\lambda)]^2},
\end{equation}
where $\gamma_\alpha=2.9$ \cite{McDonald2006}. The term $\eta_\mathrm{LSS}=7.5$ enhances the contribution of the intrinsic \lya forest variance relative to the pipeline noise, which improves the precision of the correlation function \cite{ 2023MNRAS.tmp.3626R}.

We measure the correlation functions using bins of $\Delta r =4\,h^{-1}\mathrm{Mpc}$ in both the radial and perpendicular separations. For the auto-correlations, we use 50 bins spanning separations from 0 to $200\,h^{-1}\mathrm{Mpc}$ in both directions. For the cross-correlations, we differentiate between pixels located in front of ($r_\parallel<0$) and behind ($r_\parallel>0$) quasars. With this distinction, we measure the cross-correlations using 50 bins for the perpendicular separation (0 to $200\,h^{-1}\mathrm{Mpc}$) and 100 bins for the parallel separation (-200 to $200\,h^{-1}\mathrm{Mpc}$).

We compute the covariance matrix associated with the correlations using the same approach as in previous DESI \lya analyses \cite{KP6s6-Cuceu}. First, we divide the mock survey footprint into \texttt{HEALPix} pixels \cite{Gorski2005} with $N_\mathrm{side}=16$, each covering approximately $3.7 \times 3.7 = 13.4 \,\mathrm{deg}^2$ of the sky. We then compute the correlation functions independently for each of the 1028 HEALPix pixels spanned by the DESI DR2 dataset. After that, we estimate the covariance matrix from these measurements as follows:
\begin{equation}
    C_{MN}=\frac{1}{W_M W_N}\sum_s W_M^s W_N^s (\hat{\xi}_M^s \hat{\xi}_N^s - \hat{\xi}_M \hat{\xi}_N),
\end{equation}
where $W^{XY, s}_M = \sum_{ij \in (M,s)} w_i^X w_j^Y$ represents the summed weight for a correlation in subsample $s$, $W^s = \{W^{\mathrm{AA}, s}, W^{\mathrm{AB}, s}, W^{\mathrm{QA}, s}, W^{\mathrm{QB}, s}\}$ is a vector containing the summed weights for all correlations, $\hat{\xi} = \{\xi^\mathrm{AA}, \xi^\mathrm{AB}, \xi^\mathrm{QA}, \xi^\mathrm{QB}\}$ refers to a vector containing the correlations, and $W_M = \sum_s W_M^s$ denotes the total weight.

The covariance matrix in our fiducial analysis has dimensions $15\,000 \times 15\,000$, making its estimation inherently noisy due to the limited number of subsamples. To mitigate this noise, we smooth the covariance matrix of each mock using the same procedure as in previous analyses \cite{KP6s6-Cuceu}. This smoothing method is also applied when computing the covariance of the stack of multiple mocks.

\begin{figure*}
\centering
\includegraphics[width=0.9\textwidth]{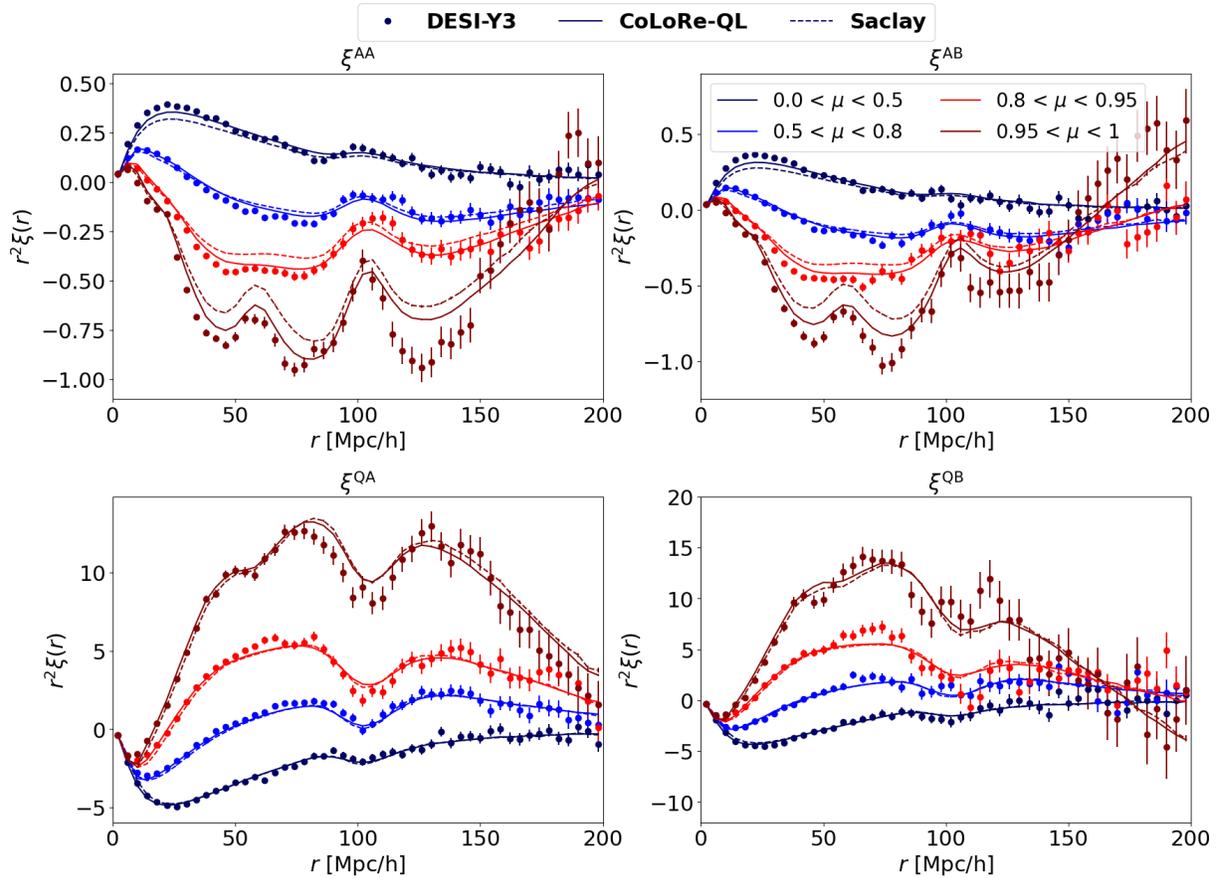}
\caption{Comparison between DESI DR2 observational measurements and mock predictions. The panels, in clockwise order, display the following: the auto-correlation of the \lyaf in region A, the auto-correlation of the \lyaf in regions A and B, the cross-correlation of the \lyaf in region B with quasar positions, and the cross-correlation of the \lyaf in region A with quasar positions. Data points and error bars represent the observational measurements and their associated uncertainties, while the solid and dashed lines represent the results from the average of 300 \QuasiLinear and 100 Saclay mocks, respectively. The colors of the lines indicate the average correlations in different wedges. Overall, the mocks show good agreement with the observational measurements; however, their accuracy diminishes for the line-of-sight wedge in the auto-correlations. 
}
\label{fig:wedges_data_mocks}
\end{figure*}

In \cref{fig:wedges_data_mocks}, we compare the four correlation functions derived from DESI DR2 observations (data points) with the averages of 300 \QuasiLinear mocks (solid lines) and 100 Saclay mocks (dashed lines). The measurements are presented as a function of radial separation, $r = (r_\parallel^2 + r_\perp^2)^{1/2}$, and the wedges defined by $\mu = r^{-1} r_\parallel$. Overall, the mock measurements agree well with the observational data across the entire range of scales, supporting the use of these mocks for assessing the performance of the correlation model (see \cref{sec:methodology_model}) in estimating the BAO position. However, we can readily see minor discrepancies between the observed correlations and the best-fit models obtained from the stack of mocks. For example, in the auto-correlations, particularly in the line-of-sight wedge, both \QuasiLinear and the Saclay mocks deviate from observed data. We attribute these differences primarily to contaminants: DLAs, redshift errors, and metal contamination, which are known to affect the shape of the correlation functions significantly. 

First, DLAs introduce a scale-dependent excess in the \lya correlation functions~\citep{FontRibera2012a, Rogers2018, Tan2025}. In our mocks we mask 75\% of the $\log{\big(\frac{N_\mathrm{HI}}{cm^{-2}}\big)}>20.3$ DLAs informed by finding algorithms applied in a mock dataset (see \Cref{app:dla_masking} and \citep{Y3.lya-s2.Brodzeller.2025}). However, the real performance in observed data is difficult to determine, thus the shape of the observed correlation functions is affected by a possible discrepancy in the assumed and the real performance. Second, redshift errors introduce spurious correlations along the line-of-sight~\citep{Youles2022, Gordon2025}, the distribution of redshift errors in observed data is undetermined, and therefore the errors introduced into our mock catalogs might not be representative of the reality. Finally, metals introduce spurious correlations along the line-of-sight (see \citep{2024arXiv240100303H} for a brief discussion). These were tuned in our mocks to reach the linear bias values found in DESI DR1~\citep{DESI2024.IV.KP6,KP6s6-Cuceu}. Although the metal biases found in DR1 and DR2 are statistically consistent they differ and therefore the amplitude of the spurious features introduced by metals might differ with the observed correlations. Finally, other differences, seen in other wedges, or in the cross-correlations can be attributed to the clustering model assumed in our mocks. \QuasiLinear mocks were created to reach bias values found in eBOSS DR12 (see \citep{Farr2020_LyaCoLoRe}), while Saclay mocks were created to match eBOSS DR16 (see \citep{Etourneau2023}), moreover the small-scale clustering of quasars in both mocks differ between each other (see~\Cref{sec:mocks}) and are yet to be refined in future realizations of mocks.

\begin{figure*}
\centering
\includegraphics[width=\textwidth]{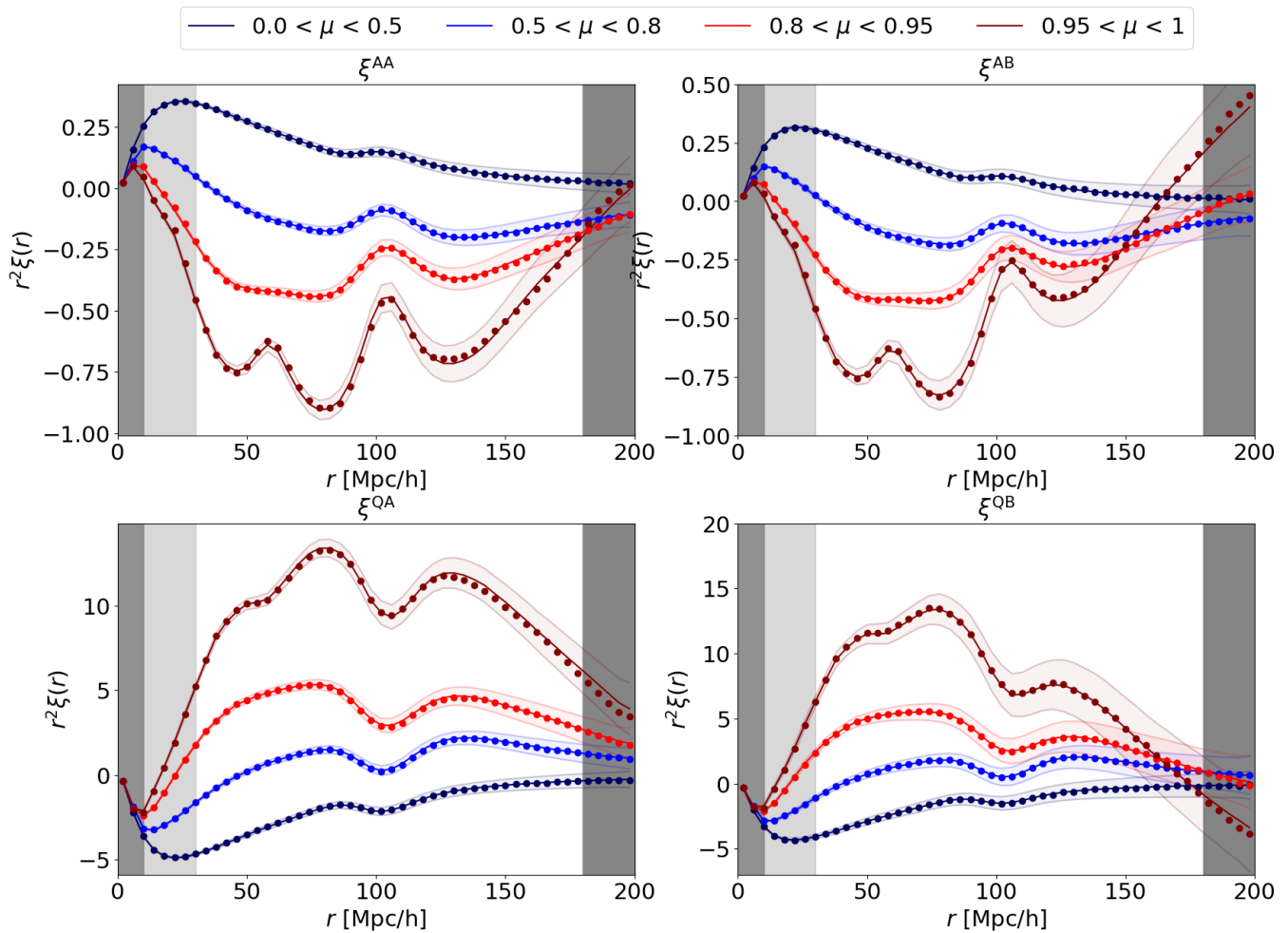}
\caption{Best-fitting model to the average correlation of 300 \QuasiLinear mocks. Dots represent the simulation measurements, lines correspond to the best-fitting model, and the shaded areas around the lines indicate the diagonal elements of the covariance matrix estimated from DESI DR2 measurements.
The dark-shaded areas highlight the scales not used in this particular fit, while the light-shaded are show the scales that are not used in the main BAO analysis of DR2.
}
\label{fig:wedges_fit_mocks}
\end{figure*}


\subsection{Modeling the correlation function}
\label{sec:methodology_model}

Our goal is to measure the BAO scale both along and across the line of sight. To this end, we use a model  
first introduced for the analysis of the Data Release 9 of BOSS \cite{Kirkby2013,Busca2013,Slosar2013}, and progressively improved over the analysis of other data releases of BOSS \cite{FontRibera2014,Delubac2015,Bautista2017,dMdB2017}, eBOSS \cite{dSA2019,Blomqvist2019,dMdB2020} and DESI \cite{2023JCAP...11..045G,KP6s6-Cuceu,DESI2024.IV.KP6}.
In what follows, we progressively build this model and detail some differences compared to DESI DR1. 

We begin with the isotropic linear matter power spectrum, $P_\mathrm{fid}$, computed evaluating \texttt{CAMB} \cite{Lewis:2019} for the same fiducial cosmology used to convert angles and redshifts into comoving coordinates in the previous section. Next, we decompose $P_\mathrm{fid}$ into an oscillatory component, $P_\mathrm{BAO}$, which contains the BAO signal, and a smooth component, $P_\mathrm{smooth}$, using the algorithm described in \cite{Kirkby2013}. To account for nonlinear effects, we apply an anisotropic Gaussian damping to the oscillatory component
\begin{align}
\label{eq:pksmooth}
P_{\mathrm{QL}}(k_\parallel,k_\perp,z)
&= P_{\mathrm{BAO}}(k,z)\,
   \exp\!\left[-\frac{k_\parallel^{2}\Sigma_\parallel^{2}
   +k_\perp^{2}\Sigma_\perp^{2}}{2}\right] \notag\\
&\quad + P_\mathrm{smooth}(k,z)
\end{align}
where $k=(k_\parallel^2 + k_\perp^2)^{1/2}$ is the wavevector, with $k_\parallel$ and $k_\perp$ as the parallel and perpendicular components, and where $\Sigma_\parallel$ and $\Sigma_\perp$ control the non-linear broadening of BAO along and perpendicular to the line of sight.

Building on the previous expression, we construct a model for the anisotropic power spectrum of the auto-correlations ($X=Y=\alpha$) and the cross-correlations ($X=\mathrm{Q}$, $Y=\alpha$) as follows:
\begin{align}
\label{eq:plya}
P_{XY}(k,\mu_k,z_\mathrm{eff})
&= b_X(z_\mathrm{eff})\,[1+\beta_X(z_\mathrm{eff})\,\mu_k^2] \notag\\
&\quad\times b_Y(z_\mathrm{eff})\,[1+\beta_Y(z_\mathrm{eff})\,\mu_k^2] \notag\\
&\quad\times G(k,\mu_k)\,F_\mathrm{NL}^{XY}(k,\mu_k)\, \notag\\
&\quad\times F_\mathrm{SM}(k,\mu_k)\,
P_\mathrm{QL}(k,\mu_k,z_\mathrm{eff}) .
\end{align}
where $\mu_k = k^{-1} k_\parallel$ denotes the cosine of the angle between the wavevector and the line of sight and the factor $b(1+\beta \mu_k^2)$, evaluated at the effective redshift of the measurement ($z_\mathrm{eff}=2.33$), accounts for the impact of linear bias and large-scale redshift-space distortions \cite{Kaiser1987}.

As discussed later in this section, the model correlation function $\xi(r_\parallel, r_\perp)$ is calculated on a
two-dimensional grid in configuration space, where each bin $r_\parallel$, $r_\perp$ has a slightly different weighted mean
redshift.  Following the procedure outlined in Section~4.2 of \cite{DESI2024.IV.KP6}, we start by defining the power spectrum at an effective redshift $z_\mathrm{eff}$ (see \cref{eq:plya}); we then compute its inverse Fourier transform to obtain the correlation function at the same effective redshift; finally, the amplitude of the correlation function is rescaled for each measured bin according to the redshift evolution of the bias parameters: $b_X(z) = b_X (z_\mathrm{eff}) \left[ (1+z) / (1+z_\mathrm{eff}) \right]^{\gamma_X}$. 
We do not include an explicit redshift evolution for $\beta$. 
Since the mean redshifts of the bins differ only slightly from
$z_\mathrm{eff}$, this approximation is well justified. 
Note that we describe the correlations involving both \lya regions (A and B) with the same model.

%

The term $G(k,\mu_k) = \mathrm{sinc}(k_\parallel \Delta r/2) \mathrm{sinc}(k_\perp \Delta r/2)$ accounts for the binning of the correlation function on a grid \cite{Bautista2017}. Following \cite{KP6s6-Cuceu}, in order to account for the finite resolution of the mocks we multiply the power spectra by a Gaussian anisotropic smoothing factor: $F_\mathrm{SM} = \exp{-[(k_\parallel \sigma_\parallel)^2 + (k_\perp \sigma_\perp)^2]/2}$.
$F_\mathrm{NL}$ accounts for non-linear corrections that are limited to relatively small scales. 
When modeling the cross-correlations, we include the terms $F_\mathrm{NL}^{\mathrm{Q}\alpha} = \exp\left[-(k_\parallel \sigma_v)^2/2\right]$ to model the combined impact of quasar redshift errors and non-linear peculiar velocities. 
When modeling the auto-correlation from observational data, we introduce a term $F_\mathrm{NL}^{\alpha\alpha}$ to account for the effects of nonlinear growth, peculiar motions, and thermal broadening on the auto-spectrum, using a model as specified by Equation 3.6 in \cite{Arinyo2015}\footnote{Since these only affect the correlations on small scales, we do not vary the parameters of this model in BAO analyses \cite{DESI.DR2.BAO.lya}.}, that accounts for non-linear growth of structure 
and thermal effects from the IGM. 
These effects are not included in the mocks, and therefore we ignore them when modeling the correlations measured from mocks.

Finally, we account for potential systematic errors in quasar redshift measurements by introducing a systematic shift ($\Delta r_\parallel$) in the line-of-sight separation between quasars and \lya absorption \cite{FontRibera2013, KP6s4-Bault}.

Next, we transform the anisotropic power spectra from Fourier to configuration space. This is done by first performing a multipole decomposition of the anisotropic power spectrum up to $\ell=6$, followed by a Hankel transform using the FFTLog algorithm \cite{Hamilton2000_fftlog} implemented in the \texttt{mcfit} package\footnote{\url{https://github.com/eelregit/mcfit}} to obtain the correlation function multipoles. From these multipoles, we then compute the two-dimensional correlation function $\xi^{XY}_M$ on a grid with spacing $\Delta r = 2\,h^{-1}\mathrm{Mpc}$. We verified that including higher-order multipoles ($\ell > 6$) does not affect BAO measurements. Additionally, we use a finer grid than the one used for measuring the correlation functions, $\Delta r = 4\,h^{-1}\mathrm{Mpc}$, to improve the precision of the model.

Next, we model the impact of astrophysical contaminants and systematics on the correlation functions. Given the range of separations considered in our analysis, we account for the influence of four metal absorption lines: SiII(1190), SiII(1193), SiIII(1207), and SiII(1260). This requires computing correlation function models for all \lya-metal, QSO-metal, and metal-metal correlations using nearly the same framework as for \lya-only correlations.
Each metal line is characterized by parameters $b_i$ and $\beta_i$, where $i$ runs over the four metal lines. Following \cite{2023JCAP...11..045G}, we fix $\beta_i=0.5$ and allow $b_i$ to vary freely in the fits. This leads to the following expressions for the auto- and cross-correlations:
\begin{equation}
    \tilde{\xi}^{\alpha \alpha} = \xi^{\alpha \alpha} + \sum_i \xi^{\alpha i} +  \sum_{ij} \xi^{ij} + \xi^\mathrm{inst},
\end{equation}
\begin{equation}
    \tilde{\xi}^{\mathrm{Q}\alpha} = \xi^{\mathrm{Q}\alpha} + \sum_i \xi^{\mathrm{Q}i} + \xi^\mathrm{TP},
\end{equation}
where $\xi^\mathrm{inst}$ and $\xi^\mathrm{TP}$ represent the contributions from DESI instrumental systematics and quasar radiation effects, respectively. However, since these effects are not included in the mocks, we chose not to account for them in the analysis (see also \cite{KP6s6-Cuceu}). 

For the auto- and cross-correlations with metals, we account for the misestimation of pixel redshifts due to the assumption that all absorption originates from \lya. Specifically, we compute the redshift of each \lya pixel as $z=\lambda/\lambda_\alpha-1$, whereas some absorptions are actually caused by metal lines, meaning their true redshifts should be $z=\lambda/\lambda_\mathrm{metal}-1$. As a result, these pixels are assigned to incorrect bins in the correlation function \cite{FontRibera2012a, Bautista2017}. 
We model this contamination with the same method used in the DR1 analysis \cite{DESI2024.IV.KP6}, with a minor modification discussed in the companion paper \cite{DESI.DR2.BAO.lya}.

As discussed in \cref{sec:methodology_forest}, we mask pixels that are most contaminated by DLAs. 
However, as discussed in \cite{Y3.lya-s2.Brodzeller.2025}, we only mask DLAs detected in spectra with relatively high signal-to-noise (SNR$>$2), where the efficiency of the DLA finders is high, finding roughly 75\% of the DLA systems\footnote{As discussed in \cref{app:dla_masking}, the impact of the exact SNR cut on the BAO parameters is minor.}.

Additionally, there are intermediate column-density systems not identified by the DLA-finding algorithm that contribute to contamination. 
Following \cite{FontRibera2012b, Rogers2018}, we account for the impact of these contaminants by adding a scale-dependence to the bias and redshift-space distortion parameters of \lya fluctuations to include the contributions from both \lya absorption and HCDs
\begin{equation}\label{eq:b_HCD}
    \tilde{b}_\alpha = b_\alpha + b_\mathrm{HCD} F_\mathrm{HCD}(k_\parallel),
\end{equation}
\begin{equation}\label{eq:beta_HCD}
    \tilde{b}_\alpha \tilde{\beta}_\alpha = b_\alpha \beta_\alpha + b_\mathrm{HCD} \beta_\mathrm{HCD} F_\mathrm{HCD}(k_\parallel),
\end{equation}
where $b_\mathrm{HCD}$ and $\beta_\mathrm{HCD}$ represent the contributions from HCD systems, while $F_\mathrm{HCD}(k_\parallel) = \exp(-L_\mathrm{HCD} \, k_\parallel)$ depends on the column density distribution of the HCDs present in the data.

The continuum fitting process (see \cref{sec:methodology_deltas}) notably alters the shape of the measured correlation functions \cite{Slosar2011,FontRibera2012b}. 
We use the formalism introduced in \cite{Bautista2017}, that multiplies the modeled correlations $\tilde \xi$ by a \textit{distortion matrix} $D^{XY}$:
\begin{equation}
    \hat{\xi}_M^{XY} = \sum_{N} D_{MN}^{XY} \xi_N^{XY},
\end{equation}
The distortion matrix can be computed from the geometry of the dataset and the distribution of weights, and in the DR2 analysis we have implemented an improved computation that takes into account the redshift evolution of the signal (see \cite{DESI.DR2.BAO.lya} for a detailed explanation and a discussion on the impact of this change).

\subsection{Fitting the correlation function} \label{sec:methodology_fitting}

By measuring the position of the BAO feature along and across the line of sight, we can constrain the distance ratios $D_M/r_d$ and $D_H/r_d$, where $r_d$ is the sound horizon at recombination, and $D_H(z)=c/H(z)$.
In order to isolate this information from the rest of the measured correlations, in BAO analyses it is common to introduce two BAO parameters ($\alpha_\perp$, $\alpha_\parallel$) that only shift the position of the peak, without affecting the smooth component of the model:
\begin{equation}\label{eq:pk_smooth}
\xi^{X Y} = \xi_\mathrm{smooth}^{X Y}(r_\parallel,r_\perp) + \xi_\mathrm{BAO}^{X Y}(\alpha_\parallel r_\parallel,\alpha_\perp r_\perp),
\end{equation}
where $\alpha_\parallel = [r^{-1}_\mathrm{d} D_\mathrm{H}(z)] / [r^{-1}_\mathrm{d} D_\mathrm{H}(z)]_\mathrm{fid}$ and $\alpha_\perp = [r^{-1}_\mathrm{d} D_\mathrm{M}(z)]/[r^{-1}_\mathrm{d} D_\mathrm{M}(z)]_\mathrm{fid}$ are defined with respect to the distances in the fiducial cosmology, denoted by ``fid".

Besides the two BAO parameters, our model to describe the correlations in mocks uses 12 nuisance parameters: 3 for \lya and quasar biases and redshift-space distortions ($b_\alpha$, $\beta_\alpha$, and $b_\mathrm{Q}$), 3 for HCD systems ($b_\mathrm{HCD}$, $\beta_\mathrm{HCD}$, and $L_\mathrm{HCD}$), 4 for the linear biases of each metal line ($b_\mathrm{SiII(1190)}$, $b_\mathrm{SiII(1193)}$, $b_\mathrm{SiIII(1207)}$, and $b_\mathrm{SiII(1260)}$), and 2 for redshift errors ($\Delta r_\parallel$ and $\sigma_v$).
We determine the best-fitting value of these parameters using \texttt{iminuit}\footnote{We verified that the difference between the best-fitting parameters and error estimates obtained from \texttt{iminuit} (\url{https://github.com/scikit-hep/iminuit}) and the PolyChord nested sampler \cite{Handley:2015a} is minimal (see also \cite{Cuceu2020}).}. 
While the BAO parameters have flat, non-informative priors, we use informative priors for some of the less constrained nuisance parameters as shown in \cref{app:nuisance}.

In the results presented in the following section, we hold fixed the value of the smoothing parameters ($\sigma_{\parallel}$, $\sigma_{\perp}$ in the $F_{\rm SM}$ term of \cref{eq:plya}) that capture numerical artifacts in the simulations (these parameters are not included in the analysis of real data).
In order to choose a value for these parameters, we do a preliminary fit of the mean of the 300 \QuasiLinear and 100 \Saclay mocks, including separations from 10 to $180~\hMpc$. We show the results from the fit in \cref{fig:wedges_fit_mocks}.
We find the values $\sigma_\parallel = 2.0~\hMpc$ and $\sigma_\perp =1.7~\hMpc$ for the \QuasiLinear mocks, and $\sigma_\parallel = 2.2~\hMpc$ and $\sigma_\perp =2.2~\hMpc$ for the \Saclay mocks.
In the case of the \QuasiLinear mocks, we also use this preliminary fit to infer the value of the non-linear BAO broadening parameters, finding $\Sigma_\perp = 3.1~\hMpc$ and $\Sigma_\parallel = 4.3~\hMpc$. These parameters are held fixed to these values in later analyses of the \QuasiLinear mocks \footnote{For the \Saclay mocks, generated from a linear power spectrum, these parameters are set to zero.}. 
While the broadening measured in the transverse direction is consistent with the theoretical expectation at this redshift ($\Sigma_\perp = 3.26~\hMpc$), the line-of-sight broadening is a bit lower than the expected one ($\Sigma_\parallel = 6.4~\hMpc$).
As discussed in \cref{ss:colore_updates}, this might be caused by the simple method that we have followed to add non-linear broadening of the BAO in the \QuasiLinear mocks.

Even though the model is able to describe fairly well the correlations on mocks down to $r=10~\hMpc$, in the rest of the analysis presented here we only fit correlations in the range of separations of $r = 30$ to $180~\hMpc$ (corresponding to 1590 and 3180 bins for the auto- and cross-correlations).
This is the configuration that was used in the DESI DR2 \lya BAO analysis \cite{DESI.DR2.BAO.lya} that we would like to validate .


\section{Results} 
\label{sec:results}

In this section, we describe and discuss the results obtained from performing the BAO analysis on the mocks. 
The mocks are used to validate the BAO analysis in two ways: in \cref{ss:stack} we use them to validate that our analysis pipeline is unbiased\footnote{\Saclay mocks were generated with a cosmology slightly different than the fiducial cosmology used in the analysis, but in terms of BAO parameters the differences are smaller than $0.03\%$}, and in \cref{sec:bao_errors} we use them to validate that the reported uncertainties on the BAO parameters are representative of the scatter between the measurement in different realizations.

\subsection{Investigating potential systematic biases in the BAO measurements}
\label{ss:stack}

\begin{figure}
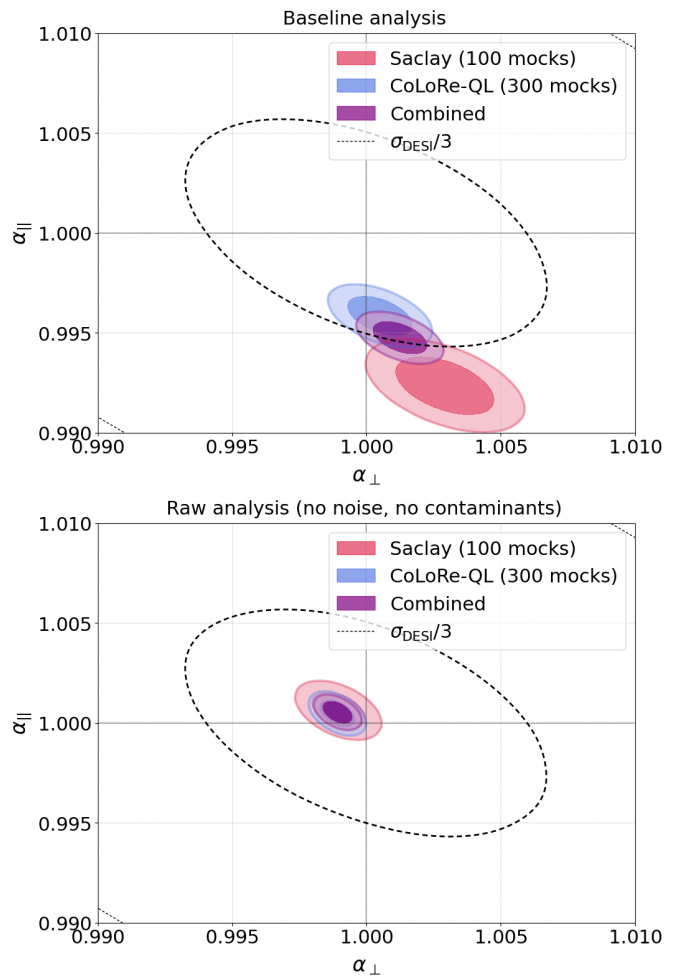

\centering
\includegraphics[width=0.49\textwidth]{stack_baseline.pdf}
\includegraphics[width=0.49\textwidth]{raw_mocks_combined.pdf}
\caption{Top panel: best fit of the stacks using the and their combination (purple). The main difference with respect to DR1 is that we do continuum fitting after adding quasar redshift errors.
As pointed out in \cite{Youles2022}, this introduces a spurious correlation that can result in a small bias on the BAO results. $\sigma_{\rm DESI/3}$ represents one third of the DR2 uncertainty. This corresponds to $0.0038$ for $\sigma_\parallel$ and $0.0045$ for $\sigma_\perp$.
Bottom panel: BAO analysis on raw mocks without instrumental noise, contaminants, or continuum fitting. Both the \texttt{Saclay} and \QuasiLinear\ mocks show no intrinsic biases, confirming that the observed shift is induced by the analysis.
}
\label{fig:stack_bao}
\end{figure}

In order to validate that our analysis pipeline produces unbiased BAO measurements, we combine the measurement from all our synthetic datasets to obtain correlation functions with very high signal-to-noise. 
Similarly to the other validation tests discussed in \cite{DESI.DR2.BAO.lya}, our requirement was to recover the true BAO parameters ($\alpha_\perp$ = $\alpha_\parallel$ = 1) within a third of the statistical uncertainty from DESI DR2.
Biases exceeding this threshold must be investigated, corrected, or incorporated into the systematic errors.

In the top panel of \cref{fig:stack_bao} we show the BAO measurements from the combined (or \textit{stacked}) correlations from 100 \Saclay mocks (red) and 300 \QuasiLinear mocks (blue), compared to the threshold of a third of the statistical uncertainty from DESI DR2 (dashed, black line).
It is clear in the figure and in \cref{tab:bao_sys_shifts}  that $\alpha_\parallel$ measurements are biased at high significance, with a bias near the threshold criterion we imposed, while $\alpha_\perp$ satisfies the criteria. 

In order to distinguish between a bias in our analysis pipeline from a bias in our simulated boxes, in the bottom panel of \cref{fig:stack_bao} we show a simplified BAO analysis where we measure the correlations directly from the \textit{raw mocks} introduced in \cref{sec:transmission}.
This analysis, detailed in \cref{app:raw_analyses}, does not require any continuum fitting and is not affected by instrumental noise or astrophysical contaminants.
The fact that these results are unbiased relative to our threshold of $\sigma_{DESI}/3$, rule out intrinsic issues with the simulated boxes.
Instead, as discussed in detail in \ref{sec:discussion}, we have identified that the bias in the baseline analysis (bottom panel) arises from the method used to incorporate redshift errors in the mocks catalogs.

\begin{table*}[]
    \centering
    \begin{tabular}{c|c|c|c|c}
         & $\Delta\alpha_{||}$ & $\Delta\alpha_\bot$ & $\Delta\alpha_{iso}$ & $\Delta\alpha_{AP}$ \\
         \hline
        \QuasiLinear\ (300) & $-0.0042 \pm 0.0006$ & $0.0005 \pm 0.0008$ & $-0.0020\pm 0.0004$ & $0.0047\pm 0.0012$ \\
         \hline
        \texttt{Saclay} (100) & $-0.0074 \pm 0.0010$ & $0.0031 \pm 0.0012$ & $-0.0027 \pm 0.0005$ & $0.0106 \pm 0.0019$ \\
         \hline
        Combined (400) & $-0.0052 \pm 0.0005$ & $0.0012 \pm 0.0006$ & $-0.0023 \pm 0.0003$ & $0.0065 \pm 0.0010$ \\
    \end{tabular}
    \caption{
    Biases on the BAO parameters ($\alpha_\parallel$, $\alpha_\perp$) from the combined analyses of 300 \QuasiLinear and 100 \Saclay mocks, as well as their combination. 
    The last two columns show the biases on the isotropic BAO parameter, defined here as $\alpha_{\rm iso} = {\alpha_\perp^{9/20} \alpha_\parallel^{11/20} }$ and the anisotropic Alcock-Paczyński parameter $ \phi = \alpha_\perp / \alpha_\parallel $.
    As discussed in \cref{sec:discussion}, these small biases can be attributed to the method used to add redshift errors in the mocks.
    }
    \label{tab:bao_sys_shifts}
\end{table*}

\subsection{Validating the BAO uncertainties}
\label{sec:bao_errors}

\begin{figure}
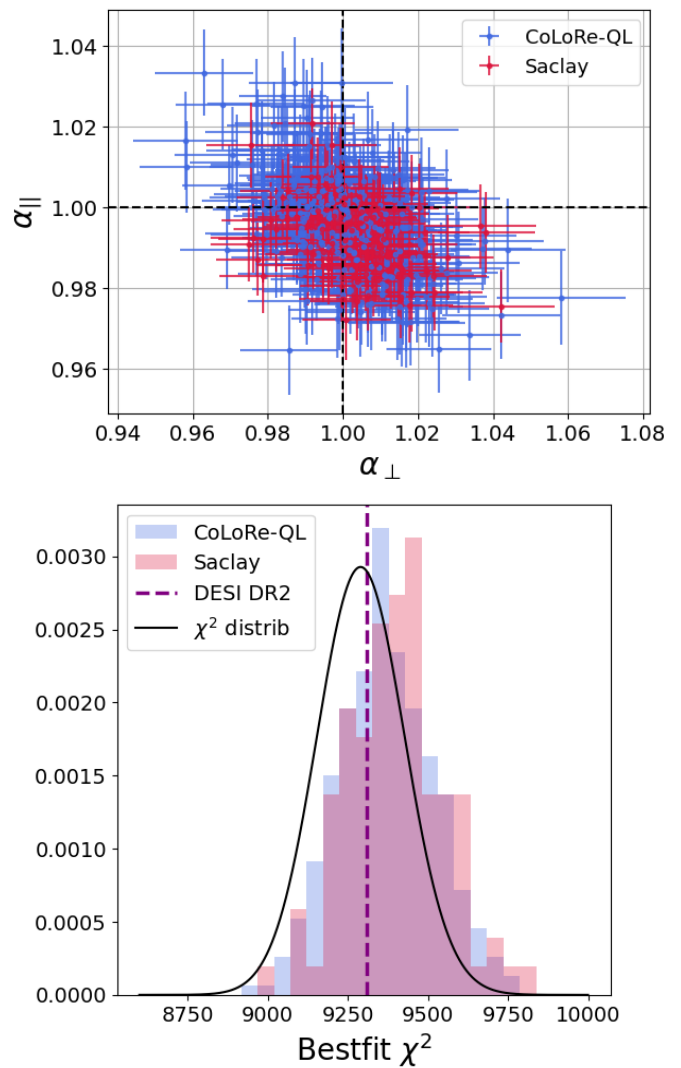

\centering
\includegraphics[width=0.5\textwidth]{mocks_scatter.pdf}
\includegraphics[width=0.45\textwidth]{mocks_chi2.pdf}
\caption{Top: Scatter plot of the best-fit BAO measurements obtained for each of the 100 \Saclay mocks (red) and 300 \QuasiLinear mocks (blue).
Bottom: Distribution of $\chi^2$ values from the fits, compared to the expected distribution given the number of degrees of freedom (black). The dashed purple line represents the $\chi^2$ of the best fit to the data.}
\label{fig:mocks_scatter_chi2}
\end{figure}

In this section we present the BAO results obtained by individually fitting each of the DESI DR2 mocks. In the top panel of Figure \ref{fig:mocks_scatter_chi2} we display the scatter of the best-fit BAO parameters across the 400 mocks (300 \QuasiLinear\ and 100 \Saclay). 
While there are no specific outliers, the measurements are not perfectly centered around the true values ($\alpha_\perp$ = $\alpha_\parallel$ = 1), particularly in the parallel direction, in agreement with the results from the stack of mocks (see \cref{tab:bao_sys_shifts}).

The bottom panel of Figure \ref{fig:mocks_scatter_chi2} shows the distribution of the best-fit \(\chi^2\) values obtained from the fits of individual mocks. 
The purple dashed line on this plot represents the $\chi^2$ of the data best-fit and the black curve represents the expected distribution based on the degrees of freedom in the fit. 
The plot shows that the distribution of $\chi^2$ values from the individual mocks is higher than expected, although the situation is significantly better than in the mocks used in the validation of DESI DR1 (see Figure 8 of \cite{KP6s6-Cuceu}). 
In \cite{KP6s6-Cuceu}, Monte Carlo simulations starting from either the best-fit model or the stacked correlation function measurements demonstrate that this shift toward higher \(\chi^2\) values arises from the model's difficulty in fitting the correlation functions,
and not from the estimates of their covariance. We have tested that this disagreement remains even when limiting the analysis to larger scales ($r>50~\hMpc$), meaning that other effects that are also present on large scales must be responsible for the somewhat large values of $\chi^2$.
Examples of these could include the impact of continuum fitting distortions \cite{Busca2025}, the anisotropy caused by DLA wings even on large separations \cite{Tan2025}, or the difficulties to model the non-standard BAO broadening in the \QuasiLinear mocks.

\begin{figure}
\centering
\includegraphics[width=0.45\textwidth]{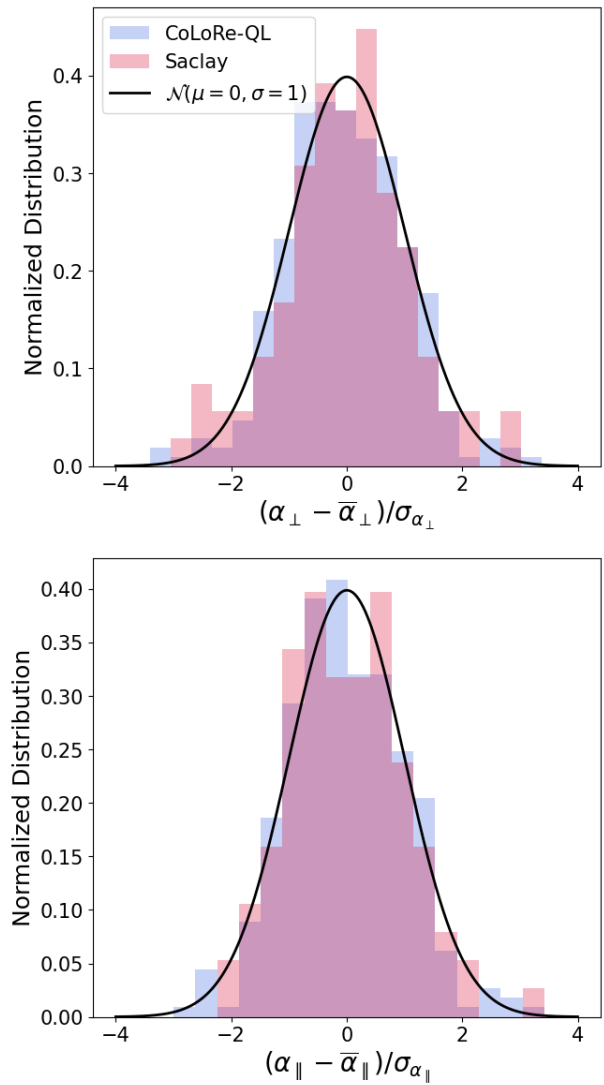}
\caption{Histograms of BAO residuals, defined as  $\Delta \alpha_{\parallel}/\sigma_{\alpha_{\parallel}}$ and $\Delta \alpha_{\perp}/\sigma_{\alpha_{\perp}}$, where the uncertainties are the Gaussian ones reported by \texttt{iminuit} in each individual fit.
The results for the 100 \Saclay mocks (red) and 300 \QuasiLinear mocks (blue) are in good agreement with a Normal distribution (black lines), indicating that the posteriors are Gaussian and the uncertainties are well estimated.}
\label{fig:mock_alphas_rms}
\end{figure}

We validate the method to estimate the BAO uncertainties in \cref{fig:mock_alphas_rms}, where we show the BAO residuals from the fit of each mock realization, defined as $\Delta\alpha_\parallel / \sigma_{\alpha_\parallel}$ and $\Delta\alpha_\perp / \sigma_{\alpha_\perp}$. The residual distributions closely follow a Normal distribution, represented by the black curve, presenting evidence that the BAO uncertainties are accurate and approximately Gaussian.

\begin{figure}
\centering
\includegraphics[width=0.45\textwidth]{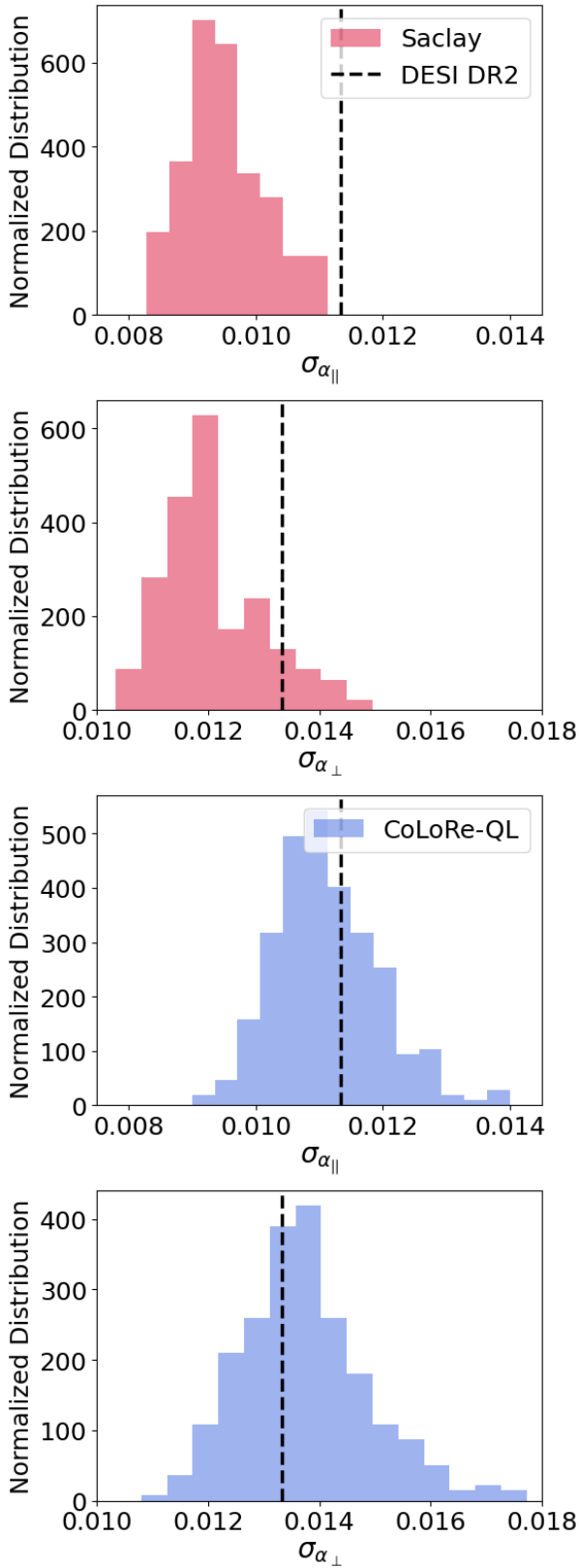}
\caption{Distribution of uncertainties on $\alpha_\parallel$ and $\alpha_\perp$ for 100 \Saclay mocks (top panel) and 300 \QuasiLinear mocks (bottom panel). The black dashed line represents the uncertainties derived from the data, which align more closely with the \QuasiLinear mocks, as the \Saclay mocks lack non-linear broadening of the BAO peak. 
}
\label{fig:sigma_alpha_mock}
\end{figure}

Finally, in \cref{fig:sigma_alpha_mock}, the uncertainties on BAO parameters of the mocks are shown alongside the statistical uncertainties from DESI data (vertical dashed line). The top panel shows the results for \Saclay mocks in red, while the bottom panel shows \QuasiLinear\ mocks in blue. As discussed in \cite{KP6s6-Cuceu}, non-linear broadening of the BAO peak modestly yet noticeably degrades the BAO measurements. Consequently, mocks that do not include this effect, such as the \Saclay mocks, exhibit uncertainties that are smaller than those observed in the data. Specifically, the uncertainties in \Saclay mocks are $10–20\%$ smaller than those in the DESI results. In contrast, the \QuasiLinear mocks demonstrate better agreement with the data, as they include a certain degree of non-linear broadening, as discussed in Section \ref{ss:colore_updates}.

\section{Discussion}
\label{sec:discussion}

In \cref{ss:stack} we presented the BAO measurements from the combined correlations measured in 300 \QuasiLinear and 100 \Saclay mocks.
As shown in the top panel of \cref{fig:stack_bao} and reported in \cref{tab:bao_sys_shifts}, we detect a small but significant bias from the expected value ($\alpha_\perp = \alpha_\parallel = 1$).
In this section we demonstrate that this bias is caused by the method used to simulate redshift errors in the mocks (\cref{ss:no_sigz}), and propose a mitigation strategy to avoid most of the contamination when analyzing real data (\cref{ss:no_close_pairs}).

\subsection{Analysis of mocks without redshift errors contamination}
\label{ss:no_sigz} 

As discussed in \cref{ss:quickquasars}, once we have already simulated the quasar spectra we add a random shift to the redshifts in our quasar catalogs, creating a mismatch between the reported redshift of the simulated quasar and the positions of the emission lines in their spectra. 
This allows us to capture the impact of redshift errors in the real catalogs, without the need to run the redshift fitter \texttt{Redrock} \cite{Redrock.Bailey.2024} for hundreds of mocks. 

The first clear impact of random quasar redshift errors in our analysis is that they smear the measured cross-correlations along the line of sight \cite{FontRibera2013}, similar to the impact of non-linear velocities (or fingers of God) in the clustering of quasars.
As discussed in \cref{sec:methodology_model}, we model this impact by multiplying the power spectrum of the cross-correlation model by $F_\mathrm{NL}^{\mathrm{Q}\alpha}=\exp\left[-(k_\parallel \sigma_v)^2/2\right]$.

Recently, \cite{Youles2022} described a second, more subtle impact of redshift errors that cause spurious correlations not only in the cross-correlation with quasars, but also in the auto-correlation. 
As discussed in \cref{sec:methodology_deltas}, in order to compute the fluctuations in the \lya forest we need to estimate the mean continua of all our quasars, as a function of rest-frame wavelength ($\bar{C}(\lambda_{\mathrm{RF}})$).
However, the random shifts added to the redshifts of our mock catalogs are translated into mis-estimations of the rest-frame wavelengths, and this causes a smoothing of the features in the mean continuum, in particular of the emission lines present in the \lya forest regions.
This effect, coupled to the clustering of the background quasars, causes spurious correlations that could bias our results.
These biases are discussed in more detail in \cite{Gordon2025}. 

\begin{figure}
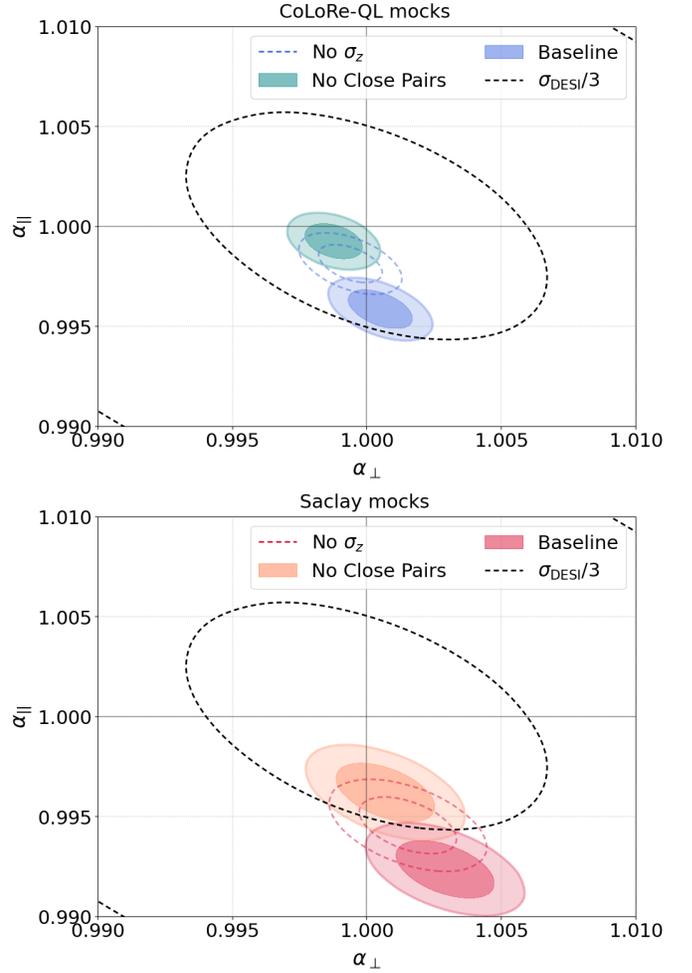

\centering
\includegraphics[width=0.49\textwidth]{zerr_comparison_colore.pdf}
\includegraphics[width=0.49\textwidth]{zerr_comparison_saclay.pdf}
\caption{Comparison of ``No Close Pairs" and "no $\sigma_z$" analyses with the baseline. Both approaches aim to mitigate contamination from redshift errors prior to continuum fitting, with ``No Close Pairs" excluding close quasar pairs (which are the most affected) and ``no $\sigma_{z}$" removing estimation redshift errors entirely before continuum fitting.}
\label{fig:comparisson_three_analyses}
\end{figure}

In order to test this hypothesis, in \cref{fig:comparisson_three_analyses} we present an alternative analysis (labeled \textit{no $\sigma_z$}) where the random redshift errors are only added \textit{after} the continuum fitting, so that they only impact the cross-correlation measurements by smoothing them along the line of sight with $F_\mathrm{NL}^{\alpha \mathrm{Q}}$.
For both sets of mocks this alternative analysis is significantly less biased than our baseline configuration, bringing it below the threshold of one third of the DESI DR2 statistical uncertainty (dashed black line ellipses).

\subsection{Mitigating the BAO scale bias by removing close pairs of quasars}
\label{ss:no_close_pairs}

Unfortunately, when analyzing real data we can not avoid having redshift errors affecting our continuum fitting. 
For this reason, in this section we propose an alternative analysis that should minimize its impact without losing a significant fraction of the data.

In a recent study, \cite{Gordon2025} presented a model that can describe the spurious correlations introduced by redshift errors. 
In that paper, the authors also show that most of the contamination in the cross-correlation comes from pixel-quasar pairs where the background quasar (whose spectrum contains the pixel) and the foreground quasar (whose position we are considering) are very close to each other. 
Similarly, they show that most of the contamination in the auto-correlation comes from pairs where one of the pixels is very close to the background quasar of the other pixel. 

Motivated by this, in \cref{fig:comparisson_three_analyses} we also show an analysis (labeled \textit{No close pairs}) where we do not include these problematic pairs, in particular those with angular separations smaller than 20 arcmin and velocity separations smaller than 4000 $\kms$.
This analysis is significantly less biased than our baseline configuration, and more importantly, it can be done when analyzing real data. 
In \cite{DESI.DR2.BAO.lya} we present the results of this alternative analysis on real data, and show that it has a negligible impact on our main BAO results. Also, to better understand the origin of the difference between the \textit{No close pairs} and \textit{no $\sigma_z$} analyses, we applied the same pair-removal cuts used in the \textit{No close pairs} analysis to the \textit{no $\sigma_z$} mocks. We found that the resulting BAO contours were essentially unchanged, indicating that these cuts do not affect the results furthermore than removing the redshift errors contamination. We therefore attribute the small differences between the analyses to statistical fluctuations.

While both alternative analyses show effectively unbiased results when analyzing the stack of 300 \QuasiLinear mocks (left panel), there seems to be a small bias when analyzing the stack of 100 \Saclay mocks, even though significantly smaller than in the baseline analysis and no longer larger than the tolerance limit of $\sigma_{DESI}/3$.
We have not been able to identify the cause for this residual bias in the analysis of the \Saclay mocks. However, we did test the Saclay mocks in both their raw (see Appendix \ref{app:raw_analyses}) and uncontaminated forms, and found no evidence of bias in either case. This suggests that the bias arises only after adding contaminants. Given that metals and HCDs are included in both the Saclay and \QuasiLinear mocks using a similar prescription, it is likely that the observed bias stems from how the mocks were constructed, rather than from the analysis framework itself.

Finally, it is important to note that the current method used to add random errors to the mocks could have exaggerated the contamination.
One of the main sources of redshift errors is that some of the quasar emission lines used to estimate their redshifts (like \ion{C}{III}, \ion{Si}{IV} or \ion{C}{IV}) can be affected by outflows or complex quasar physics. 
However, this should also affect the other high-ionization lines that are present in the \lya region,meaning the forest would shift more uniformly. This would reduce the amount of smoothing of the mean quasar continuum, the level of the spurious correlations and therefore the bias on the BAO measurements.

\section{Summary}
\label{sec:conclusions}

In this work, we use synthetic data to validate the analysis of \lya BAO measurements from the second data release (DR2) of the Dark Energy Spectroscopic Instrument (DESI), presented in \cite{DESI.DR2.BAO.lya}.
DR2 includes spectra from nearly 1.2 million quasars at redshift $z \geq 1.77$, nearly doubling the sample size of DESI DR1 \cite{DESI2024.IV.KP6}. 
As a result, DR2 achieves approximately a factor of two better statistical precision in \lya forest BAO measurements compared to DR1, thereby necessitating a more rigorous validation of the cosmological inference pipeline.

The main differences between the validation of DR2 with respect to that of DR1 (presented in \cite{KP6s6-Cuceu}) are the following.
On the one hand, we have improved the mocks.
We have increased the number of synthetic datasets used to validate the BAO analysis from 50 to 100 \Saclay mocks \cite{Etourneau2023} and from 100 to 300 \LyaCoLoRe mocks \cite{Farr2020_LyaCoLoRe}.
We have presented the \QuasiLinear mocks, an improved version of the \LyaCoLoRe mocks used in DR1 that include the non-linear broadening of the BAO peak.
On the other hand, we have improved the analysis to better mimic the analysis of real data. 
Instead of masking all the DLAs in the spectra (as done in the validation of DR1), we only mask DLAs in high signal-to-noise spectra (SNR$>$2), and only for a randomly-selected subset of 75\% to emulate the completeness of the DLA catalog used in DR2 (see the discussion in Appendix \ref{app:dla_masking}).

Using these mocks, we validated that the reported uncertainties on the BAO parameters are consistent with the scatter between the different realizations. 
Using the average measurement of correlations from all mocks, we have identified a small, but statistically significant bias in the line-of-sight BAO parameter.
This bias is close to a third of the statistical uncertainty in DR2, the threshold that we had set to consider the analysis validated.
We have shown that most of this bias is related to the redshift errors in the quasar catalog, a contamination first discussed in \cite{Youles2022}.
Following \cite{Gordon2025}, we have presented an analysis that discards a small fraction of the data that is most contaminated, and we have shown that the residual bias is significantly smaller than our statistical uncertainty. 

In the near future, we will present a cosmological analysis using the \textit{full shape} information contained in the \lya correlations, not limited to the position of the BAO peak \cite{Cuceu2021,Cuceu2023a,Cuceu2023b}. 
In order to validate these analyses, it will be important to count on improved \lya mocks that simulate the non-linear growth of structure, using perturbation theory models or similar. Non-linearities affect BAO measurements in two ways: they broaden the peak, increasing the uncertainties, and they also induce a small shift in the peak position itself, as demonstrated in recent studies \cite{deBelsunce:2024rvv,Sinigaglia:2024kub, Hadzhiyska:2025cvk}. This is discussed further in \cite{DESI.DR2.BAO.lya}, and obtaining non-linear mocks will be a key goal for future work.
It will also be important to improve the method to add redshift errors to the simulated catalogs. 
In order to do this, it will be useful to have a dedicated study of the relative redshifts of the different emission lines in the quasar spectra.

\section{Data Availability}

The data used in this analysis will be made public along the Data Release 2 (details in \cite{desi-data}). The data points corresponding to the figures from this paper will be available in a Zenodo repository \cite{laura_2025_figures}.

\acknowledgments
This material is based upon work supported by the U.S. Department of Energy (DOE), Office of Science, Office of High-Energy Physics, under Contract No. DE–AC02–05CH11231, and by the National Energy Research Scientific Computing Center, a DOE Office of Science User Facility under the same contract. Additional support for DESI was provided by the U.S. National Science Foundation (NSF), Division of Astronomical Sciences under Contract No. AST-0950945 to the NSF’s National Optical-Infrared Astronomy Research Laboratory; the Science and Technology Facilities Council of the United Kingdom; the Gordon and Betty Moore Foundation; the Heising-Simons Foundation; the French Alternative Energies and Atomic Energy Commission (CEA); the National Council of Humanities, Science and Technology of Mexico (CONAHCYT); the Ministry of Science, Innovation and Universities of Spain (MICIU/AEI/10.13039/501100011033), and by the DESI Member Institutions: \url{https://www.desi.lbl.gov/collaborating-institutions}. 

LC, JCM, AFR and ML acknowledge support from the European Union’s Horizon Europe research and innovation programme (COSMO-LYA, grant agreement 101044612). AFR acknowledges financial support from the Spanish Ministry of Science and Innovation under the Ramon y Cajal program (RYC-2018-025210) and the PGC2021-123012NB-C41 project. IFAE is partially funded by the CERCA program of the Generalitat de Catalunya. DA acknowledges support from the Beecroft Trust.

Any opinions, findings, and conclusions or recommendations expressed in this material are those of the author(s) and do not necessarily reflect the views of the U. S. National Science Foundation, the U. S. Department of Energy, or any of the listed funding agencies.

The authors are honored to be permitted to conduct scientific research on Iolkam Du’ag (Kitt Peak), a mountain with particular significance to the Tohono O’odham Nation.

\bibliographystyle{mod-apsrev4-2} 
\bibliography{DESI_supporting_papers,main}

\appendix

\section{Analyses on raw mocks}
\label{app:raw_analyses}

We refer to a BAO analysis procedure applied on the raw mocks, described in \Cref{sec:transmission}, without astrophysical contaminants, instrumental noise, or quasar continuum templates added as the \textit{raw analysis}. This analysis allows us to verify that we are able to recover the position of the BAO peak specified by the input cosmology and discard an intrinsic bias on the raw mocks.

In this analysis, we do not have to perform the astrophysical contaminants masking or the continuum fitting procedures described in \Cref{sec:methodology_forest,sec:methodology_deltas}, respectively, as they are not present in raw mocks. In other words, we compute the delta field from the transmitted flux fraction boxes directly by:
\begin{equation}\label{eq:delta_flux}
    \delta_q(\lambda) = \frac{F(\lambda)}{\overline{F}(z)} - 1.
\end{equation}

The correlation functions are computed in the same way as in the standard analysis, described in \Cref{sec:methodology_correlation}. However, in this case we do not smooth the covariance matrix since it is already positive definite given the high SNR of the raw analysis. Moreover, smoothing this covariance produces a non-positive definite matrix which induces a bias on the cross-correlation measurement on both \Saclay\ and \QuasiLinear\ mocks. 

In the case of the model, described in \Cref{sec:methodology_model}, we do not apply a distortion matrix or exclude astrophysical contaminants as these effects do not included in the raw analysis. Furthermore, we perform additional fits for the auto- and cross-correlations individually. We free the BAO parameters ($\alpha_\parallel$ and $\alpha_\perp$), the smoothing parameters ($\sigma_\parallel$ and $\sigma_\perp$), and \lya linear bias and RSD parameters ($b_\mathrm{Ly\alpha}$ and $\beta_\mathrm{Ly\alpha}$). For the cross-correlation fit, we additionally fix the quasar bias ($b_q$) value to each type of mocks truth value ($b_q = 3.123$ for \QuasiLinear\ and $b_q = 3.3$  for \Saclay).

\begin{figure}
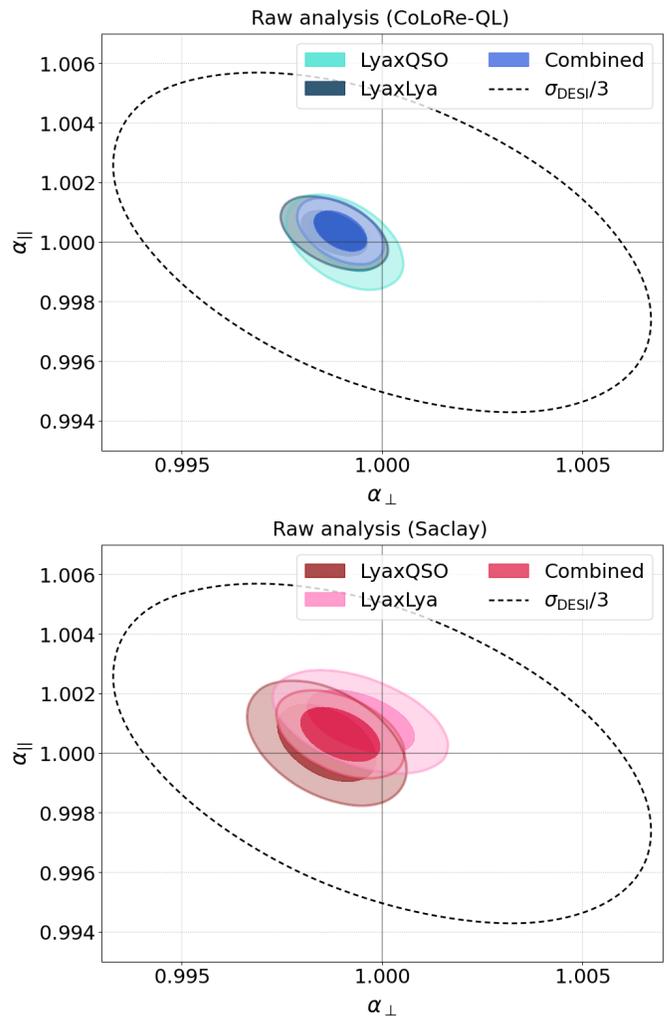

\centering
\includegraphics[width=0.49\textwidth]{raw_colore.pdf}
\includegraphics[width=0.49\textwidth]{raw_saclay.pdf}
\caption{Top panel: Fit results for the auto-correlation alone, cross-correlation alone, and combined fit for \QuasiLinear \textit{raw mocks}. In the raw analysis there is no noise, no contaminants and no continuum fitting. Bottom panel: Same fits for \Saclay \textit{raw mocks}. All fits are compatible and unbiased.}
\label{fig:raw_contours}
\end{figure}

\Cref{fig:raw_contours} shows the results for \QuasiLinear\ mocks (left panel) and \Saclay mocks (right panel). The combined fit results for $\alpha_{||}, \alpha_\bot$ are:
\begin{equation}
    \alpha_{||}^\mathrm{QL} = 1.0004 \pm 0.0005, \qquad
    \alpha_\bot^\mathrm{QL} = 0.9990 \pm 0.0004,
\end{equation}
and
\begin{equation}
    \alpha_{||}^\mathrm{Saclay} = 1.0009 \pm 0.0006, \qquad
    \alpha_\bot^\mathrm{Saclay} = 0.9992 \pm 0.0007,
\end{equation}
for \QuasiLinear\ and \Saclay mocks, respectively. All the results presented in \Cref{fig:raw_contours} are within a third of the statistical uncertainty from DESI DR2 threshold and close to the $\alpha_{||}=\alpha_{\bot}=1$ target value, confirming that the raw mocks are suitable to be used in the analysis validation presented in this work.

\section{Impact of DLA masking} 
\label{app:dla_masking}

Damped Lyman-$\alpha$ absorbers (DLAs) are found and masked in the BAO analysis to optimize the statistical precision of the measurement, and any residual DLA contamination is accounted for via free parameters in the model fit. 
In this section, we use mocks to study the impact on BAO results when changing the DLA-finding algorithm and to find the optimal DLA catalog configuration for the DR2 BAO analysis on real data.

The \lya BAO analysis from DESI DR1 \cite{DESI2024.IV.KP6} masked those DLAs that were found by both a Gaussian Process (GP) finder \citep{Ho:2020DLAFinderGP} and a convolutional neural network (CNN) finder \citep{Wang2022}. 
Tests on one mock found the combination to produce unbiased BAO results compared to a catalog of all input DLAs~\citep{KP6s6-Cuceu}. 
In preparation for DR2 analysis, several improvements have been made to the GP and CNN finders, and an additional finder (referred to as the template, or TMP, finder) has been developed that fits each spectrum with a template for quasars and one or more DLA models \citep{Y3.lya-s2.Brodzeller.2025}.

To test the performance of these finders for the DR2 BAO analysis, we ran all three finders on one realization of the \texttt{LyaCoLoRe} mocks\footnote{This mock was generated before the improvements described in \cref{ss:colore_updates} and does not have non-linear broadeding of the BAO.}.
After each DLA finder was run on the mock data, several quality cuts were made to the raw output catalogs. 
In a first series of analyses, all three finders were reduced to only include DLAs found in spectra for which the mean signal-to-noise measured on the red-side of the \lya forest was $SNR>3$. 
Furthermore, only DLAs with high column density, $\log{\big(\frac{N_\mathrm{HI}}{cm^{-2}}\big)}>20.3$ for the $N_\mathrm{HI}$ calculated by each algorithm were kept. 
We included additional quality cuts, for the CNN confidence parameter \texttt{CONF}$>0.5$; for the GP probability parameter \texttt{P\_DLA}$>0.9$; and for the TMP enforcement of \texttt{DLAFLAG}=0 (see \citep{Wang2022, Ho:2020DLAFinderGP, Y3.lya-s2.Brodzeller.2025} for further details on these parameters). 
The performance of the different algorithms, in terms of purity and completeness of their DLA catalogs, is discussed in the companion paper \cite{Y3.lya-s2.Brodzeller.2025}, and we refer the interested reader to that publication for more detail.
Here we complement this study by looking at the impact on the BAO parameters when using different DLA finders.

\begin{figure}
    \centering
    \includegraphics[width=.9\linewidth]{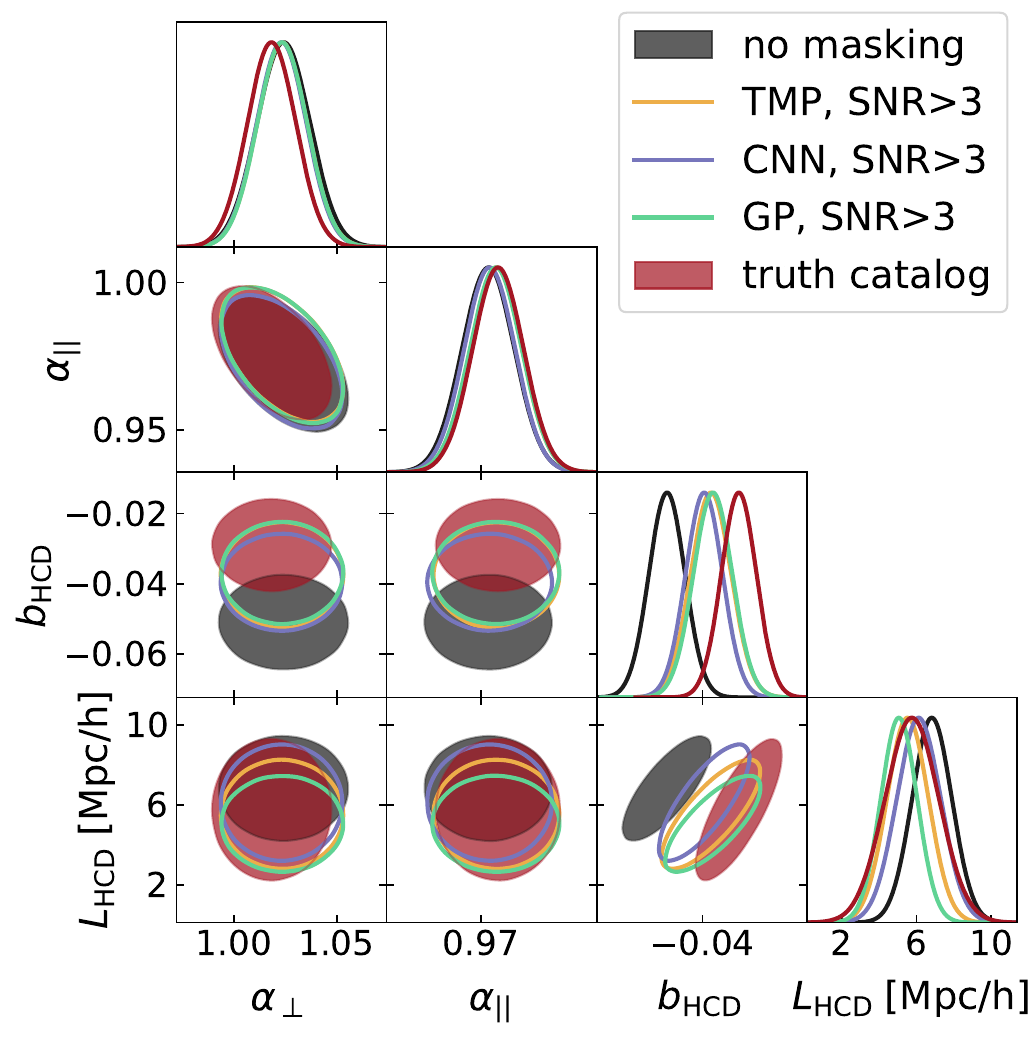}
    \caption{Contours for HCD parameters and BAO parameters under different DLA masking schemes. Only 95\% confidence levels are shown. The results from the three DLA finders are shown in open contours to facilitate visualization, while the two extremes (\textit{no masking} and \textit{truth catalog}) are shaded. For $b_\mathrm{HCD}$, the finder-based analyses produce values lying between the two extremes as expected, and the $L_\mathrm{HCD}$ varies depending on the typical length scale masked. 
    However, the BAO parameters (upper left) are stable to variations. 
    The BAO parameters are in mild tension with the true input cosmology values because this is an analysis on only a single mock; the stack of mocks is unbiased as discussed in \Cref{ss:stack}.}
    \label{fig:contours_dla_testing}
\end{figure}

We ran different end-to-end BAO analyses on this particular \LyaCoLoRe mock, but using different catalogs to mask DLAs, including an analysis without masking any DLA (labeled \textit{no masking}) and an analysis that masked all the DLAs added to the mocks (labeled \textit{truth catalog}).
\Cref{fig:contours_dla_testing} shows that the BAO parameters are very robust to the DLA catalog used, and the small differences can be explained by statistical fluctuations.
However, each analysis has a different amount of residual contamination from DLAs that were not identified, and this translates into different values for the nuisance parameters describing the contamination by High Column Density systems (HCDs): $L_\mathrm{HCD}$, $\beta_\mathrm{HCD}$, and $b_\mathrm{HCD}$, presented in \Cref{sec:methodology_model}.
As expected, $b_\mathrm{HCD}$ varies between a large negative value for the \textit{no masking} case, where a large contribution to the bias due to DLAs must be taken into account in the model, to a small negative value when all DLAs are removed and only small HCD contributions remains (\textit{truth catalog} masking). 
The three DLA finders result in very similar values for $b_\mathrm{HCD}$ and all lie between the two extremes, having found most but not all DLAs in the fiducial catalog. $L_\mathrm{HCD}$, the typical length scale of DLA systems that appear in the data, is large when none are masked (because even large DLAs remain in the spectra) and decreases with any form of masking, because only smaller systems remain. The model succeeds in capturing the correlation function variations through the HCD parameters such that the choice of DLA masking does not bias the BAO parameters, as evidenced by the stability in $\alpha_\parallel$ and $\alpha_\perp$ parameters. 

In order to decide on the optimal method to mask DLAs in the DR2 analysis, \cite{Y3.lya-s2.Brodzeller.2025} studied possible ways to combine the catalogs constructed by the different DLA finders.
Combined catalogs containing DLAs detected by at least two finders, when limited to spectra with $SNR>2$, resulted in similar performances in terms of completeness and purity of the catalogs, both around 75\%.
The recommendation from \cite{Y3.lya-s2.Brodzeller.2025}, that was adopted in the \lya BAO measurement from DR2 of \cite{DESI.DR2.BAO.lya}, was to mask DLAs that were identified by the GP finder and one of the other two finders \footnote{We use the values of redshift and column density reported by the GP finder.}, a combination that we label as \textit{GP+}. 

\begin{figure}
    \centering
    \includegraphics[width=.95\linewidth]{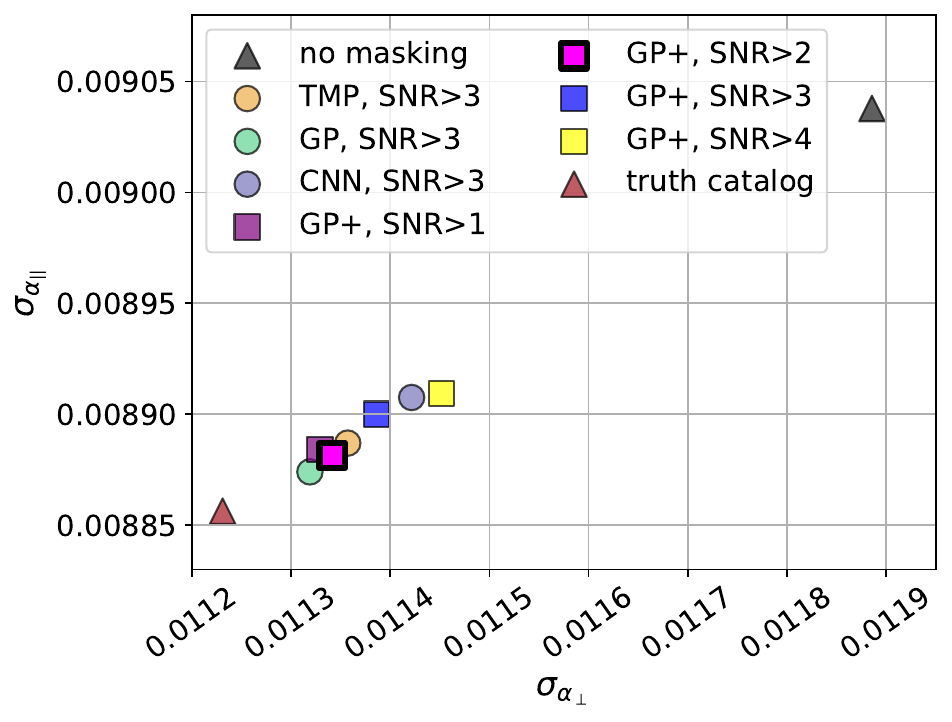}
    \caption{Forecasted uncertainties on $\alpha_\parallel$ and $\alpha_\perp$ for the one-finder (circles) and two-finder (squares) DLA catalogs compared to the case of no masking and masking all objects in the truth catalog (triangles). The final choice for data analysis is the GP+, SNR$>2$ catalog (pink square with bold borders), which has an optimal completeness-purity tradeoff in addition to performing with high precision, as shown here.
    }
    \label{fig:dla_forecast}
\end{figure}

After the decision was made to use the \textit{GP+} catalog, we revisited the impact of different SNR cuts. 
Masking DLAs detected in lower SNR spectra would trade a lower purity of the sample for a higher completeness. 
In \cref{fig:dla_forecast} we compare the forecasted BAO uncertainties for different DLA masking strategies, in particular when varying the SNR threshold in the GP+ catalog.
However, as the different finders mask different \lya data, there is some natural sample variation in both the signal and the uncertainty of the resulting BAO parameters. 
This makes it challenging to directly compare the BAO uncertainties to make the optimal choice. 
To more robustly quantify the performance of each masking catalogue, we make a BAO forecast using \texttt{vega}. The forecast works by creating a vector that represents the correlation function but is purely equal to the (noiseless) best-fit model for that analysis. Using the covariance matrix of the mock data in combination with that vector, the BAO fit is performed, obtaining BAO uncertainties that are not affected by statistical fluctuations. 

\cref{fig:dla_forecast} shows that the error on $\alpha_\parallel$ and $\alpha_\perp$ is highest when no DLAs are masked, and moves lower for the various finders, approaching the \textit{truth catalog} case with smallest errorbars. Although GP on its own results in the smallest errors of the single-finder cases (circles), as previously stated, we deemed it more robust to require detection in at least one of the other finders. Of the SNR cuts tested for the GP+ catalog, the SNR$>2$ and SNR$>1$ versions perform best, so we choose SNR$>2$ for its higher purity for the BAO DR2 data analysis. 

Given the trend to lower $\sigma_{\alpha_\perp}$ and $\sigma_{\alpha_\parallel}$ for increasing numbers of masked DLAs, it is interesting to question whether masking increasingly smaller DLAs would continue to reduce the error bars. Given that masking also removes \lya signal, however, there must be a point at which the gains in precision due to removal of these contaminants are outweighed by the loss in signal. 
We explored this idea by masking not only all the DLAs added to the mocks, but also all the HCD systems with increasingly lower column density. 
By running \texttt{vega} forecasts, we find that the uncertainties continue to decrease until a threshold of $\sim\log{(N_\mathrm{HI})}>19$. 
However, beyond this, continuing to mask systems with even lower column density causes the uncertainties to increase again due to excess loss in \lya signal.
This investigation is not relevant to the DESI DR2 analysis, as the DLA finders are not expected to perform well at such low column densities; however, it demonstrates that with future improved algorithms, the analysis could benefit from masking lower column density systems.

\section{Details on the nuisance parameters}
\label{app:nuisance}

As discussed in \cref{sec:methodology_fitting,sec:results}, we perform various types of analyses on the mocks presented throughout this work. \Cref{tab:priors} provides a description of the free parameters used to model the correlation functions from our mocks in the range $30<r<180\ h^{-1}\mathrm{Mpc}$, with the corresponding priors listed in the second column. 

The last two rows of this table specify the priors used for the finite-grid Gaussian smoothing parameters ($\sigma_\parallel,\sigma_\perp$), and the non-linear BAO peak broadening parameters ($\Sigma_\parallel,\Sigma_\perp$). These priors were applied exclusively in a fit on the high-precision BAO measurements described in \cref{ss:stack} over $10<r<180~\hMpc$ to determine these parameters values. For all other analyses, these parameters remain fixed at $\sigma_\parallel = 2.0~\hMpc$, $\sigma_\perp = 1.7~\hMpc$, $\Sigma_\parallel = 4.3\ h^{-1}\mathrm{Mpc}$, and $\Sigma_\perp = 3.1\ h^{-1}\mathrm{Mpc}$ for \QuasiLinear\ mocks, while for \Saclay mocks, they are fixed at $\sigma_\parallel = \sigma_\perp = 2.2~\hMpc$ and $\Sigma_\parallel = \Sigma_\perp = 0$.

For the raw analyses presented in \cref{app:raw_analyses}, we only allow the BAO scale parameters ($\alpha_\parallel,\alpha_\perp$), the \lya linear bias and RSD parameter ($b_\alpha,\beta_\alpha$), the quasar linear bias ($b_{\rm q}$), the finite-grid Gaussian smoothing parameters ($\sigma_\parallel,\sigma_\perp$), and the non-linear BAO peak broadening parameters ($\Sigma_\parallel,\Sigma_\perp$) to vary. All other parameters are excluded from the model in this specific type of analyses.

\begin{table*}[!htbp]
\centering
\caption{Free parameters and priors used in the $30<r<180~\hMpc$ model fitting presented in this work. $U(\mathrm{min},\mathrm{max})$ denotes flat priors in the $[\mathrm{min},\mathrm{max}]$ range. $\mathcal{N}(\mu,\sigma)$ denotes Gaussian priors of mean $\mu$ and $\sigma$ standard deviation. The priors set to the parameters on the last two rows are only used in a $10<r<180~\hMpc$ fit of the correlation functions to determine these parameters values and remain fixed for all other analyses.}
\label{tab:priors}

\begin{tabular}{llp{9cm}}
    \multicolumn{1}{c}{\textbf{Parameter}} & \multicolumn{1}{c}{\textbf{Prior}} & \multicolumn{1}{c}{\textbf{Description}} \\
  \toprule
    $\alpha_\parallel,\alpha_\perp$                   & $U(0.01,2.0)$          & BAO scale position parameters. \\
    $b_{Ly\alpha}$                                    & $U(-2.0,0.0)$          & Lyman-$\alpha$ linear bias. \\
    $\beta_{Ly\alpha}$                                & $U(0.0,5.0)$           & Lyman-$\alpha$ RSD parameter. \\
    $b_{\rm q}$                                       & $U(0.0,10.0)$          & Quasar linear bias. \\
    $\sigma_v\,[h^{-1}\mathrm{Mpc}]$                  & $U(0.0,15.0)$          & Statistical QSO redshift estimate errors amplitude. \\
    $\Delta r_{\rm shift}\,[h^{-1}\mathrm{Mpc}]$      & $\mathcal{N}(0.0,1.0)$ & Systematic QSO redshift estimate shift. \\
    $b_{\rm HCD}$                                     & $U(-0.2,0.0)$          & High Column Density (HCD) systems linear bias. \\
    $\beta_{HCD}$                                     & $\mathcal{N}(0.5,0.09)$ & HCD RSD parameter. \\
    $L_{\rm HCD}$                                     & $\mathcal{N}(5.0,1.0)$  & Typical length scale of unmasked HCDs. \\
    $b_{\rm \ion{Si}{II}(1190)}$                      & $U(-0.5,0.5)$         & Linear bias of the \ion{Si}{II}(1190) transition. \\
    $b_{\rm \ion{Si}{II}(1193)}$                      & $U(-0.5,0.5)$         & Linear bias of the \ion{Si}{II}(1193) transition. \\
    $b_{\rm \ion{Si}{III}(1207)}$                     & $U(-0.5,0.5)$         & Linear bias of the \ion{Si}{III}(1207) transition. \\
    $b_{\rm \ion{Si}{II}(1260)}$                      & $U(-0.5,0.5)$         & Linear bias of the \ion{Si}{II}(1260) transition. \\
    \hline
    $\sigma_\parallel,\sigma_\perp\,[h^{-1}\mathrm{Mpc}]$ & $U(0.0,10.0)$         & Finite-grid Gaussian smoothing parameters. \\
    $\Sigma_\parallel,\Sigma_\perp\,[h^{-1}\mathrm{Mpc}]$ & $U(0.0,20.0)$         & Non-linear broadening of the BAO peak parameters. \\
\end{tabular}
\end{table*}

\end{document}